\documentclass[final,5p,times,twocolumn]{elsarticle}

\usepackage{amssymb}
\usepackage{lipsum}
\usepackage{booktabs}
\usepackage{multirow}
\usepackage{booktabs}
\usepackage{amsmath}
\usepackage{graphicx}
\usepackage{subcaption}
\usepackage{algorithm}
\usepackage{algorithmicx}
\usepackage{algpseudocode}
\usepackage{tikz}

\journal{Physics Letters B}

\begin{document}

\begin{frontmatter}



\title{ECG Identity Authentication in Open-set with Multi-model Pretraining and Self-constraint Center \& Irrelevant Sample Repulsion Learning}


\author[first]{Mingyu Dong}
\author[first]{Zhidong Zhao\textsuperscript{*,}}
\author[first]{Hao Wang}
\author[first]{Yefei Zhang}
\author[first]{Yanjun Deng}
\affiliation[first]{organization={Hanghzou Dianzi University},
            addressline={}, 
            city={Hangzhou},
            postcode={310018}, 
            state={Zhejiang},
            country={China}}





\begin{abstract}
Electrocardiogram (ECG) signal exhibits inherent uniqueness, making it a promising biometric modality for identity authentication. As a result, ECG authentication has gained increasing attention in recent years. However, most existing methods focus primarily on improving authentication accuracy within closed-set settings, with limited research addressing the challenges posed by open-set scenarios. In real-world applications, identity authentication systems often encounter a substantial amount of unseen data, leading to potential security vulnerabilities and performance degradation.
To address this issue, we propose a robust ECG identity authentication system that maintains high performance even in open-set settings. Firstly, we employ a multi-modal pretraining framework, where ECG signals are paired with textual reports derived from their corresponding fiducial features to enhance the representational capacity of the signal encoder. During fine-tuning, we introduce Self-constraint Center Learning and Irrelevant Sample Repulsion Learning to constrain the feature distribution, ensuring that the encoded representations exhibit clear decision boundaries for classification.
Our method achieves 99.83\% authentication accuracy and maintains a False Accept Rate as low as 5.39\% in the presence of open-set samples. Furthermore, across various open-set ratios, our method demonstrates exceptional stability, maintaining an Open-set Classification Rate above 95\%.
\end{abstract}



\begin{keyword}
Electrocardiogram \sep Identity Authentication \sep Open Set \sep Multi-modal



\end{keyword}

\end{frontmatter}




\section{Introduction}
\label{introduction}
\thanks{*Corresponding author}
Electrocardiogram (ECG) signals capture the electrical activity of the heart throughout its cardiac cycle, with each individual exhibiting unique physiological characteristics. Due to these inherent individual differences \cite{hoekema2001geometrical}, ECG signals have gained increasing attention in recent years as a biometric modality for identity authentication \cite{sumalatha2024deep}. Compared to well-established biometric technologies such as facial recognition and fingerprint identification \cite{uwaechia2021comprehensive}, ECG authentication offers distinct advantages in terms of security and resilience against spoofing. The intrinsic uniqueness of ECG signals, coupled with their dynamic nature, makes them significantly more difficult to replicate or forge, positioning ECG as a promising and robust feature for biometric identification. Advancements in signal acquisition technology have significantly enhanced the quality of non-invasive data collection, enabling the capture of clear and well-defined waveforms \cite{pereira2023biometric}. This progress has not only improved the accuracy and reliability of acquired signals but has also facilitated seamless and unobtrusive acquisition methods. As a result, the feasibility and practicality of utilizing such signals for identity authentication have been greatly enhanced, offering a more user-friendly and efficient approach to biometric verification.

In early research on ECG identity authentication, fiducial features were introduced to determine morphological characteristics that could be used for identity differentiation \cite{poree2016ecg}. A commonly employed category of these features includes the time intervals between standard medical fiducial points on the ECG waveform, such as P, Q, R, S, and T waves. In addition to these fiducial features, various non-fiducial features have been explored. Examples of such features include principal components \cite{irvine2008eigenpulse}, wavelet coefficients \cite{chan2008wavelet}, and autocorrelation coefficients \cite{wang2007analysis}. After extracting these features, identity authentication is performed by comparing the similarity between the extracted features and the stored reference templates in the database. This similarity assessment determines whether the input belongs to the same individual.

With the advancement of deep learning, the application of deep learning models for ECG signal classification has gradually replaced traditional identity authentication methods based on direct similarity comparison. Deep learning models, composed of multiple hidden layers, learn sample distributions through non-linear mappings and ultimately classify the extracted features. Both methods \cite{wang2024ecg} and \cite{krishnamoorthy2024deep} utilize deep learning-based approaches, while method \cite{aslan2024visgin} employs graph neural networks for identity verification. Additionally, Transformer \cite{jin2025reading} and Mamba \cite{qiang2024ecgmamba} models have also been explored for ECG abnormal classification, further expanding the range of deep learning techniques applied in this domain.

\begin{figure*}[h!] 
\centering 
\includegraphics[width=\textwidth]{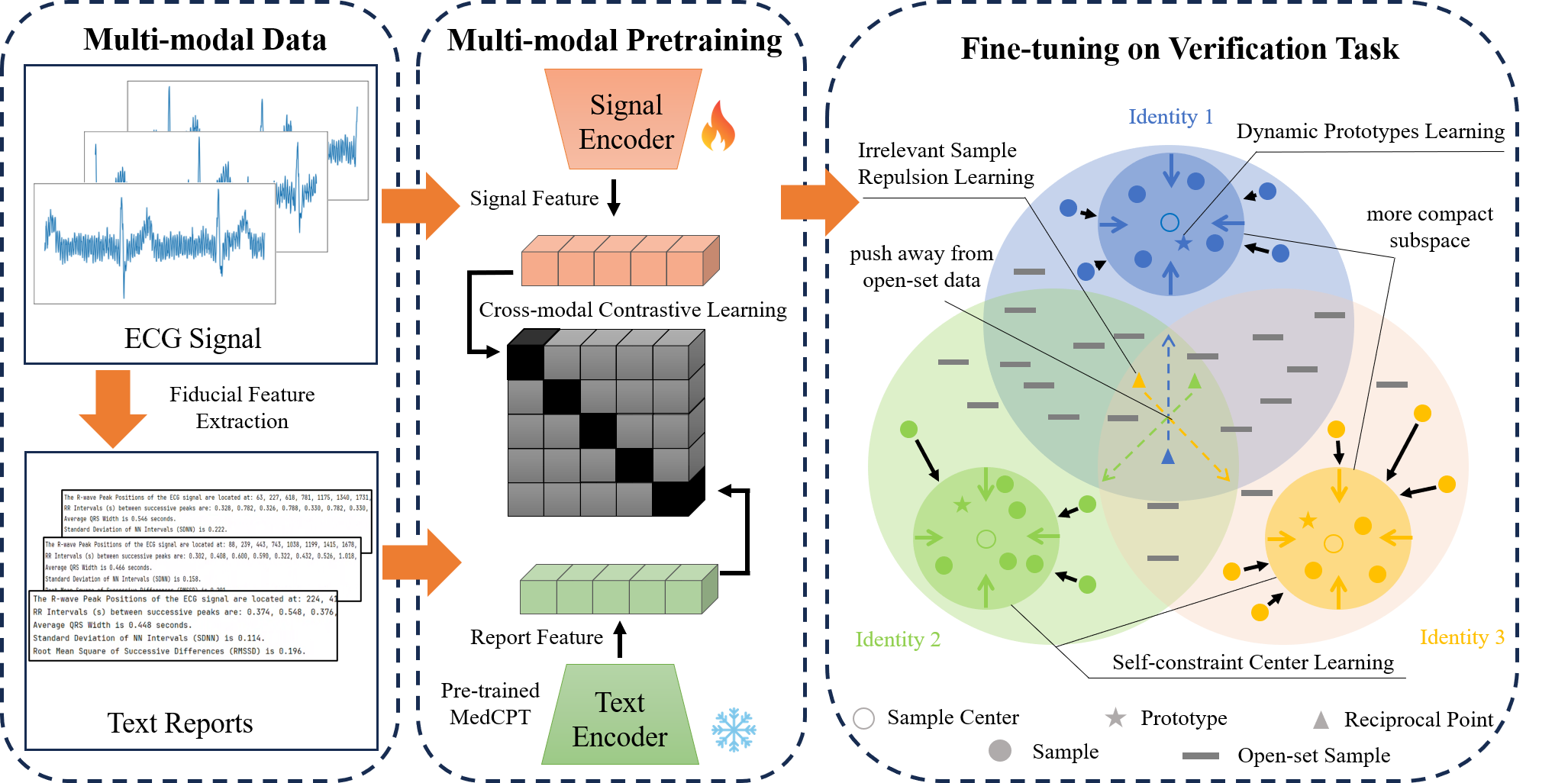} 
\caption{The proposed method is outlined in the workflow diagram, which consists of two main components: multi-modal pretraining and fine-tuning on identity authentication task. Within component B, we introduce Self-constraint Center Learning and Irrelevant Sample Repulsion Learning.} 
\label{fig:flowchart} 
\end{figure*}

In the field of ECG disease diagnosis, Liu et al. proposed a dual-modal approach that integrates both ECG signals and clinical reports for disease classification \cite{liu2024zero}. This multi-modal framework leverages the complementary nature of textual and signal modalities, demonstrating that incorporating clinical text into the training process can significantly enhance the model’s ability to represent ECG signals. However, in the domain of identity authentication, textual data has rarely been utilized as an additional modality in training pipelines. Given that textual information often encapsulates crucial contextual knowledge, its integration could play a vital role in improving the capability of signal encoders to capture identity-related features within ECG data. By incorporating text as an auxiliary modality, identity authentication models can achieve more robust and discriminative representations.

In real-world identity authentication scenarios, systems often face saturation attacks, where they struggle to effectively distinguish whether an input sample belongs to a registered class. Most identity authentication systems based on classification models assign a predefined label to each input signal, making them inherently vulnerable to security risks. Wu et al. were the first to highlight the issue of system stability in ECG identity authentication under open-set conditions \cite{wu2021scalable}. Building upon this foundation, this study conducts a comprehensive robustness evaluation of ECG identity authentication across a broader range of open-set data.

To enhance the capacity for ECG signal representation of the model, we propose a multi-modal pretraining module that integrates both ECG signals and fiducial feature text report. During the fine-tuning phase, we introduce a novel training strategy combining Self-constraint Center Learning and Irrelevant Sample Repulsion Learning. These modules enable the model to achieve better classification capability even without additional open-set data, improving the generalization to unseen open-set samples. Our method effectively reshapes the distribution of sample features, ensuring that training samples are mapped into a more compact and well-structured subspace. Meanwhile, features extracted from open-set samples, which are not present during training, are constrained within a predefined range, preventing them from interfering with the labeled feature space. In smaller-scale open-set scenarios, our method effectively distinguishes between known and unknown samples with high precision. This capability ensures that the model not only accurately determines whether an input sample belongs to a previously registered identity but also reliably classifies and verifies registered identities. By enhancing the model’s ability to differentiate between in-distribution and out-of-distribution samples, our method strengthens the robustness of identity authentication systems. The contributions of this paper are as follows:




\begin{itemize}
  \item \textbf{Multi-modal pretraining.} We leverage a large dataset to pre-train a multi-modal model that integrates ECG signals with fiducial feature text report. Using contrastive learning, we align the modalities, ensuring a correspondence between the information in the text and the ECG signal.
  
  \item \textbf{Self-constraint Center Learning.} 
   The sample center is defined as the centroid of the sample feature distribution. During training, sample features are explicitly encouraged to move closer to their respective class centers. To mitigate potential biases arising from imbalanced training data distributions, we further incorporate a Dynamic Prototype Learning mechanism. This component adaptively adjusts the class prototypes throughout training, promoting a more balanced and representative feature distribution across classes.

  \item \textbf{Irrelevant Sample Repulsion Learning.}To approximate the distribution of open-set data without explicitly involving any open-set samples during training, we introduce the concept of irrelevant sample. These samples are designed to represent the distributional characteristics of all other identity classes beyond the target identity class, within the constraints of a limited dataset. By encouraging target identity class samples to be distant from these irrelevant samples, the model is guided to form more distinct and well-separated decision boundaries. 
\end{itemize}

\section{Methods}

This section focuses on our method of ECG identity authentication in an open-set setting. In Section \ref{problem}, we first define the problem formally. The subsequent sections detail our proposed module, which comprises three key components: Multi-modal pretraining, Self-constraint Center Learning, and Irrelevant Sample Repulsion Learning. The overall workflow is illustrated in Figure \ref{fig:flowchart}.

\subsection{Problem Definition}
\label{problem}
Given a set of ECG signals with their identities $\mathcal{D}_{L}=\{(x_{1},id_{1}),...,((x_{n},id_{n})\}$. $N$ is the registered identities $id_{i}\in \{1,...,N\}$, $id_{i}$ is the identity of the signal $x_{i}$. Given another larger amount of test data \( \mathcal{D}_T = \{t_1, \dots, t_u\} \) where the identity of \( t_i \) belongs to \( \{1, \dots, N\} \cup \{N+1, \dots, N+U\} \) which contains close-set data and open-set data. The \( U \) is the number of unregistered identities in realistic scenarios. the deep embedding feature of category \( k \) is denoted by \( S_k \) and $S_k \in \mathcal{D}_{L}$. 



For the model $\psi$, it is crucial to learn the feature representations of registered users from the training dataset $\mathcal{D}_{L}$ and establish a well-defined feature distribution. During the training phase, the model must establish clear decision boundaries between samples of different identity labels while ensuring sufficient separation $R$ between these boundaries. To enhance the generalization to open-set scenarios despite a limited number of training samples, all samples except those belonging to a designated label $id$ are treated as open-set data $\mathcal{D}_L^{\neq id}$. During the testing phase, the dataset includes open-set samples, introducing unseen categories that were not present during training. The model is designed to achieve two \textbf{primary objectives}: (1) accurately classify samples belonging to previously registered identities $id \in \{1, \dots, N\}$, ensuring high recognition performance within known identities; and (2) effectively identify and exclude unregistered samples $id \in \{N+1, \dots, N+U\}$, minimizing false acceptances of unseen identities.

\subsection{Multi-modal Pretraining}
\label{pretrain}

In ECG signals, fiducial features refer to key points or waveform characteristics with significant physiological relevance \cite{boumbarov2009ecg}. Prior research has demonstrated their practicality and effectiveness in ECG-based identity authentication systems \cite{wu2020ecg, lee2018efficient, plataniotis2006ecg}. In this work, we aim to leverage fiducial features to enhance the representational capacity of the ECG signal encoder by aligning waveform structures with physiologically meaningful cues.

To this end, we select five representative fiducial features that capture both the morphological and dynamic aspects of ECG signals: R-wave peak positions, RR intervals, QRS widths, standard deviation of NN intervals (SDNN), and root mean square of successive differences (RMSSD). R-wave peak positions, RR intervals, and QRS widths characterize the overall morphology of the ECG waveform, while the latter two (SDNN and RMSSD) quantify signal variability, which also reflects individual-specific cardiac patterns. These extracted fiducial features are then transformed into a structured textual report to facilitate interpretability and downstream processing. The report follows a predefined template format denoted as \textit{‘The R-wave Peak Positions of the ECG signal are located at: \{\}’. RR Intervals between successive peaks are: \{\}. Average QRS Width is \{\} seconds. Standard Deviation of NN Intervals is \{\}.Root Mean Square of Successive Differences is \{\}.}, enabling consistent alignment between waveform features and their corresponding semantic representations.

There is a corresponding relationship between ECG signals and text reports. Extracting relevant information from texts to assist model training benefits the identity authentication. Therefore, in this section, we leverage both ECG signals and their associated text reports to pre-train a model capable of effectively capturing and representing ECG signal features. This pre-trained model serves as a robust feature extractor, facilitating the downstream identity authentication task.

$\{(s_{1}, r_{1}),(s_{2}, r_{2}),...,(s_{i}, r_{i})\}$ donated as the ECG signals with the corresponding text reports. To independently extract features from the input data pair, separate encoders are employed for different modalities. Specifically, the report encoder $\mathcal{F}_{r}$ utilizes a pre-trained model and tokenizer from MedCPT \cite{jin2023medcpt}, while the signal encoder $\mathcal{F}_{s}$ adopts various model configurations to process the ECG signals. Following feature extraction, the non-linear projection layers $\mathcal{P}_{e}$ and $\mathcal{P}_{s}$ are applied to unify the feature dimensions across modalities, where the features are extracted as $z_{r,i}=\mathcal{P}_{r}(\mathcal{F}_{r}(r_{i}))$ and $z_{s,i}=\mathcal{P}_{s}(\mathcal{F}_{s}(r_{i}))$. To align the representations from different modalities, contrastive learning is employed. The cosine distances between the two modalities can be denoted as $s^{s2r}_{i,i}=z_{s,i}^\top z_{r,i}$ and $s^{r2s}_{i,i}=z_{r,i}^\top z_{s,i}$, respectively. These values serve as quantitative measures of the similarity between the extracted feature representations from each modality, providing insights into their alignment and compatibility within the learned feature space. The loss function during the pretraining process can be formulated as follows:
\begin{equation}
\mathcal{L}_{i,j}^{s2r} = -\log \frac{\exp(s_{i,j}^{s2r}/\tau)}{\sum_{k=1}^L \mathbb{I}_{[k\neq i]} \exp(s_{i,k}^{s2r}/\tau)},
\end{equation}

\begin{equation}
\mathcal{L}_{i,j}^{r2s} = -\log \frac{\exp(s_{i,j}^{r2s}/\tau)}{\sum_{k=1}^L \mathbb{I}_{[k\neq i]} \exp(s_{i,k}^{s2r}/\tau)},
\end{equation}

\begin{equation}
\mathcal{L}_{\text{Contrastive}} = \frac{1}{2L} \sum_{i=1}^N \sum_{j=1}^N \left( \mathcal{L}_{i,j}^{r2s} + \mathcal{L}_{i,j}^{r2s} \right).
\end{equation}
where, $\mathcal{L}_{i,j}^{s2r}$ and $\mathcal{L}_{i,j}^{r2s}$ represent the signal-report and report-signal cross-modal contrastive losses, respectively. The temperature hyper-parameter, denoted as $\tau$, is set as 0.07 in the experiment. $L$ is the batch size per step, which is a subset of $N$.

\subsection{Self-constraint Center Learning}

The pre-trained model obtained through the aforementioned procedure demonstrates strong representational capability for ECG signal. In the subsequent stage, the pre-trained model is fine-tuned on the target dataset for identity registration. For the training dataset $\mathcal{D}_{L}$, the model is required to correctly assign labels in the classification task. 

To address this issue, this paper introduces the concept of sample self-constraining, which aims to map the feature representations of training samples into a more compact subspace. By constraining the feature distribution, the model can better distinguish from open-set samples. Specifically, we define $C$ as the centroid of the sample distribution, ensuring that feature representations remain clustered around a meaningful reference point, which improves both intra-class compactness and inter-class separability. For each identity category $id$, all samples belonging to the same identity in the $\mathcal{D}_{L}$ are collected. We define the class-specific sample center $C$ by identifying the instance with the minimum total distance to all other samples within the same identity $id$. Specifically, for each sample, we compute the sum of pairwise distances to all other samples in the identity class, and select the one with the smallest cumulative distance as the representative center:
\begin{equation}
    C_{id} = \arg \min_{m \in S^{d}} \sum_{i=1}^{n} \sum_{j \neq i}^{n} d(m_{id}^{i},m_{id}^{j})
\end{equation}
here, $m$ denotes the signal latent feature extracted by the signal encoder, while $n$ represents the number of samples belonging to a given identity $id$. The pairwise distance between samples is denoted by $d(a, b)=\sqrt{(a-b)^2}$. These centers serve as key reference points for analyzing the distribution and structural characteristics of the identity representations.

During the training process, each sample in the training set is constrained by the \( L_2 \)-norm to regulate the distance between its feature representation and the corresponding sample center $C$. The loss function is formulated as follows:
\begin{equation}
\mathcal{L}_{self}(x;\theta) = \frac{1}{M \cdot N} \sum_{id=1}^{M} \sum_{i=1}^{N} \left\| \mathcal{F}_{r}(x_{i}^{id}) - C_{id} \right\|_2^2
\end{equation}
where $C_{id}$ represents the sample center corresponding to the specific identity label $id$. While training, the sample centers are dynamically updated at the beginning of each epoch, ensuring centers adapt to the evolving feature distributions.

The softmax function is commonly used in classification tasks to map feature representations to probability distributions, serving as the basis for loss computation. However, empirical analysis reveals that features transformed by softmax tend to exhibit a loosely distributed structure \cite{dhamija2018reducing}. Therefore, in this section, we propose an alternative approach that replaces the traditional softmax-based classification loss with a distance-based metric, leveraging pairwise sample distances to enhance feature compactness.

Inspired by \cite{yang2018robust}, we introduce the concept of dynamic prototypes. Unlike center $C$, which are typically fixed at the centroid of sample feature distributions, dynamic prototypes $P^{k}$ are learnable parameters that adaptively update the positions based on the distance between the given samples and the prototypes. Sample center $C$, when estimated from a limited number of samples, is susceptible to shifts influenced by the distribution of outlier or atypical samples. In contrast, a dynamic prototype can effectively represent the learned distribution of a given identity by continuously adapting to the encoding space of the model. By leveraging prototype learning, we provide a dual safeguard against potential biases caused by shifts in the true distribution center. This dynamic adjustment mechanism enables samples to achieve a more balanced distribution between the sample center $C$ and the dynamic prototypes $P$. In the method, the probability of features in the softmax function is replaced with the distance between samples and their respective prototypes $P$:
\begin{equation}
\label{prob}
Prob(x \in P^{id}|x) = \frac{e^{- d(\mathcal{F}_{r}(x^{id}),P^{id})}}{\sum_{k=1}^M e^{- d(\mathcal{F}_{r}(x^{k}),P^{k})}}
\end{equation}
where $d(\mathcal{F}_{r}(x^{id}),P^{id})=\left\| \mathcal{F}_{r}(x^{id}) - P^{id} \right\|_2^2$, $M$ is the number of identity labels. The probability in the cross-entropy loss is represented by Equation \ref{prob}. During the training process, we also use the distance $d(, \cdot ,)$ to update the parameters of the prototypes. As a result, the overall loss function is defined as the sum of the modified cross-entropy loss and the prototype update loss:
\begin{equation}
\mathcal{L}_{proto}(x;\theta,P) = \frac{1}{M \cdot N} \sum_{id=1}^{M} 
 \sum_{i=1}^{N} (-\log(Prob(x_{i}))+ d(\mathcal{F}_{r}(x^{id}_{i}),P^{id}))
\end{equation}

\subsection{Irrelevant Sample Repulsion Learning}
After applying self-constrained center learning and dynamic prototype learning, the sample feature distribution is confined to a more compact subspace. However, in the absence of additional open-set datasets, effectively distinguishing whether a given sample belongs to a registered identity remains a critical challenge. Inspired by adversarial reciprocal points learning \cite{chen2021adversarial}, we propose irrelevant sample repulsion learning. The feature representation of the reciprocal points $\mathcal{P}^{id}$ can be formulated as $\mathcal{D}_L^{\neq {id}} \cup \mathcal{D}_U$. The reciprocal points $\mathcal{P}^{id}$ of identity label $id$ should be as close as possible to the feature set of non-$id$ dataset $\mathcal{D}_L^{\neq id}$ and the open dataset $\mathcal{D}_U$. 
\begin{equation}
\max \left( \zeta \left( \mathcal{D}_L^{\neq {id}} \cup \mathcal{D}_U, \mathcal{P}^{id} \right) \right) \leq R.
\end{equation}
both $\mathcal{P}^{id}$ and $R$ are learnable parameters. By imposing a constraint on the maximum distance between the $\mathcal{P}^{id}$ and the sample features, achieving the separation of registered samples from those in the open set. 

In real-world scenarios, the training process does not involve open-set data which is often characterized by an almost infinite amount of samples and categories. Given the limited amount of training data with identity labels, where features from the open-set data and labeled data are complementary, we shift the training objective from the open-set data to the labeled data. This shift allows us to effectively leverage the complementary nature of these two types of features. The corresponding loss function is expressed as follows:
\begin{equation}
{\cal L}_o(x;\theta,{\cal P}^k,R^k) = \frac{1}{M \cdot N} \sum_{id=1}^{M} 
 \sum_{i=1}^{N}\max(d_e( \mathcal{F}_{r}(x_i),{\cal P}^{id}) - R^{id}, 0),
\end{equation}
where $d_e( \mathcal{F}_{s}(x),{\cal P}^k)=\frac{1}{N} \left\| \mathcal{F}_{r}(x_i) - {\cal P}^{id} \right\|_2^2$. By imposing the \( L_2 \)-norm constraint on the distance $R^{id}$ between the training samples and the $\mathcal{P}^k$, ensuring that the samples $\mathcal{D}_L^{= {id}}$ are distanced from those of other identity labels $\mathcal{D}_L^{\neq {id}}$.

\begin{algorithm}
\caption{SimCLR’s main learning algorithm.}
\label{alg:simclr}
\begin{algorithmic}[1]
\Require batch size $N$, num of batch $L$, sample $x \in \mathcal{D}_{L}$, dynamic prototype $P$, reciprocal point $\mathcal{P}$ , learnable margin $R$, structure of encoder $\mathcal{F}_{s}$


\For{all $k \in \{1, \dots, N\}$} 
    \State $S_{k}=\mathcal{F}_{s}(x_{id})$
    \State $C_{k}=\arg \min \sum_{i=1}^{n} \sum_{j \neq i}^{n} d(S_{id}^{i},S_{id}^{j})$
\EndFor

\For{sampled minibatch $\{x_k\}_{k=1}^{N}$} 
    \For{all $k \in \{1, \dots, M\}$} 
        \State $S_{k}=\mathcal{F}_{s}(x_{k})$
        \State $d_{self}^{k}=d_{e}(S_{id}^{k}, C_{id})$
        \State $d_{proto}^{k}=d_{e}(S_{id}^{k}, P^{id})$
        \State $d_{reciprocal}^{k}=d_{e}(S_{id}^{k}, \mathcal{P}^{id})$
    \EndFor
    \State $\mathcal{L}_{self}= \frac{1}{L} \displaystyle\sum^{L} d_{self}$
    \State $Prob(x \in P^{id}|x) = \frac{e^{-d_{proto}^{k}}}{\sum^K e^{-d_{proto}^{K}}}$
    \State $\mathcal{L}_{proto} = \frac{1}{L} \displaystyle\sum^{L}(-\log(Prob(id=k|x)) + d_{proto}^{k})$
    \State ${\cal L}_o = \frac{1}{L} \displaystyle\sum^{L} (\max(d_{reciprocal}^{k} - R, 0))$
    \State $\mathcal{L} = \alpha \mathcal{L}_{self} + \beta \mathcal{L}_{proto} + \gamma \mathcal{L}_{O}$
    \State update $\mathcal{F}_{s}$, $P$, $\mathcal{P}$ and $R$ to minimize $\mathcal{L}$
\EndFor
\State \textbf{return} encoder network $\mathcal{F}_{s}(\cdot)$
\end{algorithmic}
\end{algorithm}

\subsection{Training Process of the Proposed Method}

The overall training procedure is outlined in Algorithm \ref{alg:simclr}. The ECG encoder $\mathcal{F}_{s}$ utilized in this process is the multimodal pre-trained model from Section \ref{pretrain}. Before entering the fine-tuning loop, we compute the class centers $C$ for all identity labels. These centers are dynamically updated in each training iteration to ensure the most accurate representation of their locations. The distance between each sample and its corresponding center $d_{e}(x, C)$ is used as a self-constraint center loss, encouraging samples to move closer to their respective class centers. This constraint helps refine the feature representations by minimizing the discrepancy between individual samples and their centers.

Once inside the loop, the distance between each sample and all dynamic prototypes $P$ is calculated, replacing traditional probability-based computation with a distance-based approach. During prototype loss computation, we incorporate the sample-to-prototype distance $D_{e}(x, P)$, encouraging samples to move closer to their respective prototypes. Additionally, we compute the distance between samples and designated reciprocal points $d_{e}(x, \mathcal{P})$, enforcing separation by penalizing deviations from a learnable margin $R$. Finally, the three loss components are weighted and summed to the overall loss $\mathcal{L}$, updating the parameters of $\mathcal{F}_{s}$, $P$, $\mathcal{P}$ and $R$.

\begin{table*}[h!]
\caption{
A comparative study of ECG signal encoders with different backbone architectures. The ResNet-based encoders include ResNet18, ResNet34, and ResNet50, while the ViT-based encoders consist of ViT-Tiny, ViT-Small, and ViT-Base. }
\label{bigtable}
\centering
\setlength{\tabcolsep}{3pt} 
\begin{tabular}{l c|| c c c c c || c c c }
\toprule
\multicolumn{1}{c}{\multirow{2}{*}{dataset}} & \multirow{2}{*}{backbone} & \multicolumn{5}{c||}{Close-set evaluation} & \multicolumn{3}{c}{Open-set evaluation} \\  \cline{3-10}
& & ACC[\%]  & f1 score[\%]  & Precision[\%]  & Recall[\%]  & AUC[\%] & OSCR[\%] &  FAR[\%]  &  TNR[\%]\\ 
\midrule
\multirow{6}{*}{ECGID} 
 & ResNet18 & 96.11 & 98.71 & 96.80 & 96.11 & 99.96 & 89.24 & 5.39 & 49.03  \\
 & ResNet34 & 96.67 & 98.52 & 97.17 & 96.67 & 99.96 & 89.79 & 5.35 & 54.67  \\
 & ResNet50 & 95.00 & 97.89 & 95.80 & 95.00 & 99.95 & 89.03 & 5.37 & 52.67  \\ 
 & ViT tiny & 68.89 & 76.38 & 70.81 & 68.89 & 95.33 & 53.45 & 9.62 & 38.45  \\
 & ViT small & 62.78 & 71.24 & 65.06 & 62.78 & 92.31 & 49.43 & 9.93 & 37.03 \\
 & ViT base & 41.7 & 46.16 & 42.03 & 41.67 & 91.50 & 26.40 & 12.98 & 42.96  \\
 \midrule
\multirow{6}{*}{MITBIH} 
 & ResNet18 & 99.60 & 99.79 & 99.61 & 99.60 & 99.99 & 97.60 & 7.53 & 53.70 \\
 & ResNet34 & 99.43 & 99.79 & 99.45 & 99.43 & 99.99 & 95.16 & 8.09 & 55.69 \\
 & ResNet50 & 99.50 & 99.75 & 99.51 & 99.50 & 99.97 & 94.20 & 8.32 & 53.88 \\ 
 & ViT tiny & 90.63 & 92.44 & 90.98 & 90.63 & 99.20 & 80.04 & 10.69 & 42.92 \\
 & ViT small & 86.80 & 88.84 & 87.02 & 86.80 & 98.71 & 72.47 & 11.98 & 41.79 \\
 & ViT base & 66.23 & 70.31 & 68.99 & 66.23 & 95.75 & 54.62 & 14.08 & 39.96 \\
  \midrule
\multirow{6}{*}{Autonomic} 
 & ResNet18 & 98.40 & 98.79 & 97.93 & 98.40 & 99.83 & 95.84 & 6.21 & 52.31 \\
 & ResNet34 & 98.52 & 98.84 & 98.04 & 95.52 & 99.90 & 94.40 & 6.56 & 29.55  \\
 & ResNet50 & 96.52 & 97.75 & 96.13 & 96.52 & 99.81 & 91.45 & 7.11 & 21.32   \\ 
 & ViT tiny & 94.52 & 96.83 & 94.73 & 94.52 & 99.74 & 87.01 & 7.86 & 46.33    \\
 & ViT small & 93.20 & 95.57 & 93.43 & 93.20 & 99.61 & 83.56 & 8.09 & 30.64    \\
 & ViT base & 74.00 & 75.97 & 75.11 & 74.00 & 99.23 & 58.64 & 11.67 & 35.17  \\
\bottomrule
\end{tabular}
\label{backbone}
\end{table*}

\section{Results and discussions}

\subsection{Implementation Details}
In our framework, the multimodal pretraining phase leverages the MIMIC-ECG dataset \cite{gow2023mimic}, which comprises 800,035 pairs of signal-text data. Each signal is recorded at a 500 Hz sampling rate for a duration of 10 seconds. Due to computational constraints, a subset of 100,000 high-quality samples was selected for pretraining in our experiments. For the ECG signal encoder, we employed both ResNet1D \cite{koonce2021resnet} and Vision Transformer (ViT) \cite{han2022survey} models with varying hyperparameters to explore the impact of architectural choices on performance. For the text encoder, we utilized the pretrained MedCPT model \cite{jin2023medcpt}, which is specifically tailored for medical text processing. The pretraining process was conducted over 50 epochs using the AdamW optimizer \cite{kinga2015method}. All experiments were executed on an NVIDIA 4070 GPU.

\textit{Dataset}: During the fine-tuning phase for identity authentication, the datasetss were sourced from the following three repositories:

\textbf{ECGID} \cite{lugovaya2005biometric} contains 310 ECG recordings, obtained from 90 persons, digitized at 500 Hz with 12-bit resolution over a nominal ±10 mV range. The records were obtained from volunteers (44 men and 46 women aged from 13 to 75 years who were students, colleagues, and friends of the author). 

\textbf{MIT-BIH Arrhythmia} \cite{moody2001impact} Database contains 48 half-hour excerpts of two-channel ambulatory ECG recordings, obtained from 47 subjects studied by the BIH Arrhythmia Laboratory. The recordings were digitized at 360 Hz.

\textbf{Autonomic Aging} \cite{schumann2022autonomic} which collects the high-resolution biological signals to describe the effect of healthy aging on cardiovascular regulation. The ECG data in Autonomic are recorded from 1121 healthy volunteers, which contains two different collection modes. The sampling rate of the ECG signal is 1000Hz and the length is longer than 8 minutes.

\textit{Metrics}: In the experiments, two categories of evaluation metrics were employed: \textbf{Closed-set} and \textbf{Open-set} metrics. \textbf{Closed-set metrics} assess the authentication accuracy for enrolled users, focusing on scenarios where all test samples belong to known classes. We utilized a comprehensive set of metrics, including Accuracy (ACC), F1 Score, Precision, Recall, and Area Under the Curve (AUC). 

\textbf{Open-set metrics} were designed to evaluate performance in more realistic scenarios. The test dataset contains a large number of signals from unregistered users. The Open Set Classification Rate (OSCR) is utilized to evaluate the model's balanced discrimination capability between known and unknown identity categories \cite{dhamija2018reducing}. A threshold $\delta$ is employed to determine whether the unknown identity samples belong to the registered categories. The calculation of OSCR involves two key metrics. The Correct Classification Rate (CCR) measures the proportion of samples that are correctly classified:
\begin{equation}
CCR(\delta) = \frac{|\{x \in D_T^k \wedge \arg\max_k Prob(k|x) = k \wedge Prob(\hat{k}|x) \geq \delta\}|}{|D_T^k|}.
\label{eq:ccr}
\end{equation}
The False Positive Rate (FPR) assesses the proportion of unknown samples that are incorrectly classified as registered users:
\begin{equation}
FPR(\delta) = \frac{|\{x | x \in D_U \wedge \max_k Prob(k|x) \geq \delta\}|}{|D_U|}.
\label{eq:fpr}
\end{equation}
The OSCR is defined as the area under the curves of the CCR and the FPR across varying $\delta$.

In addition, we chose the False Accept Rate (FAR) and True Negative Rate (TNR) as supplementary evaluation metrics. FAR measures the proportion of unregistered samples that are incorrectly authenticated as valid users, highlighting the vulnerability to false acceptance. TNR evaluates the proportion of unregistered samples correctly identified as unknown, providing insight into the robustness in distinguishing genuine from unauthorized inputs.
\begin{equation}
FAR(\delta) = \frac{|\{x | x \in D_U \wedge \max_k Prob(k|x) \geq \delta\}|}{|D_L|+|D_U|}.
\end{equation}

\subsection{Experiment of backbone of ECG Encoder}

\begin{figure*}[htbp]
    \centering
    \begin{subfigure}[b]{0.33\textwidth}
        \includegraphics[width=1.1\textwidth]{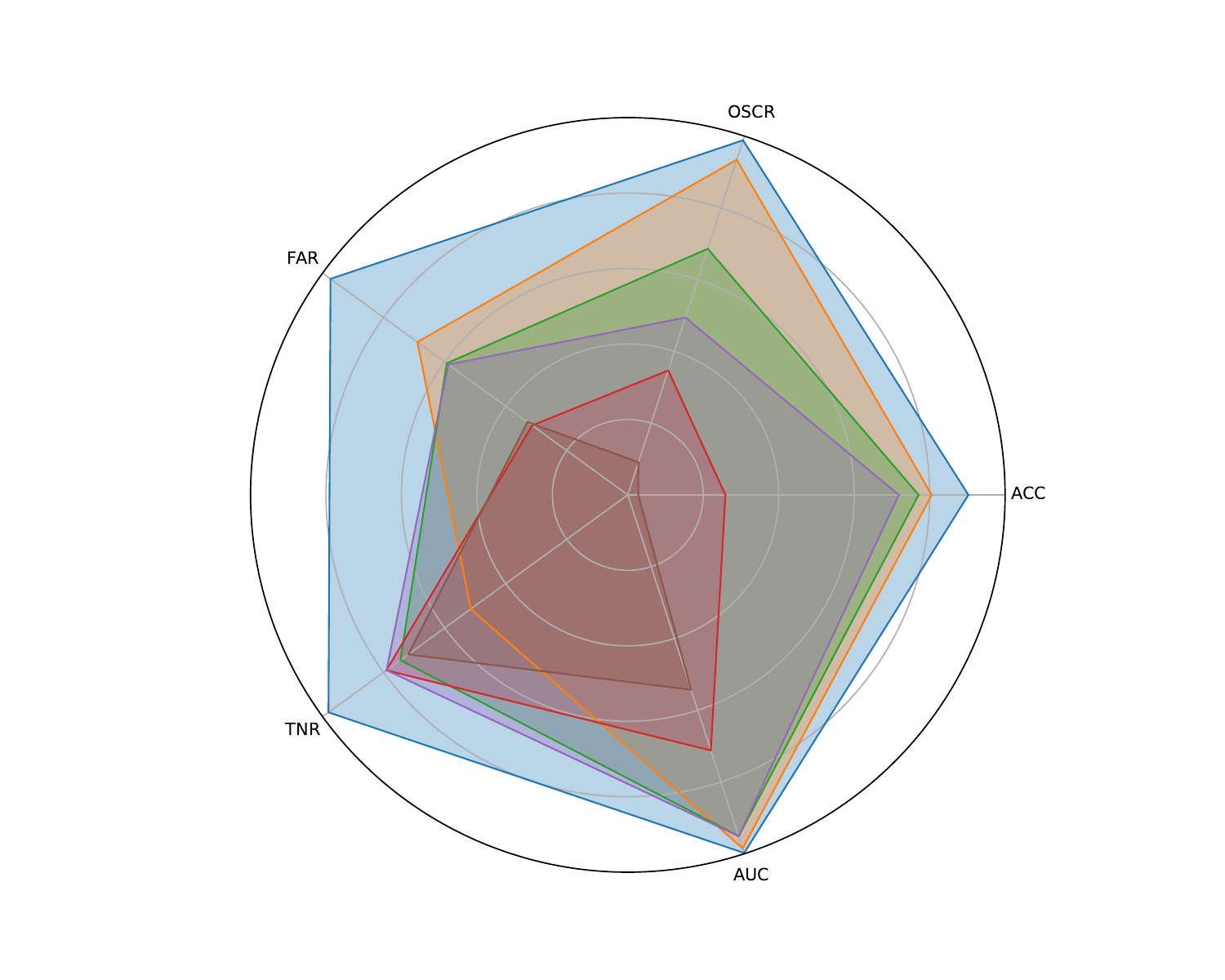}
        \caption{Experiment with ECGID dataset}
        \label{fig:subfig1}
    \end{subfigure}
    \hfill
    \begin{subfigure}[b]{0.33\textwidth}
        \includegraphics[width=1.1\textwidth]{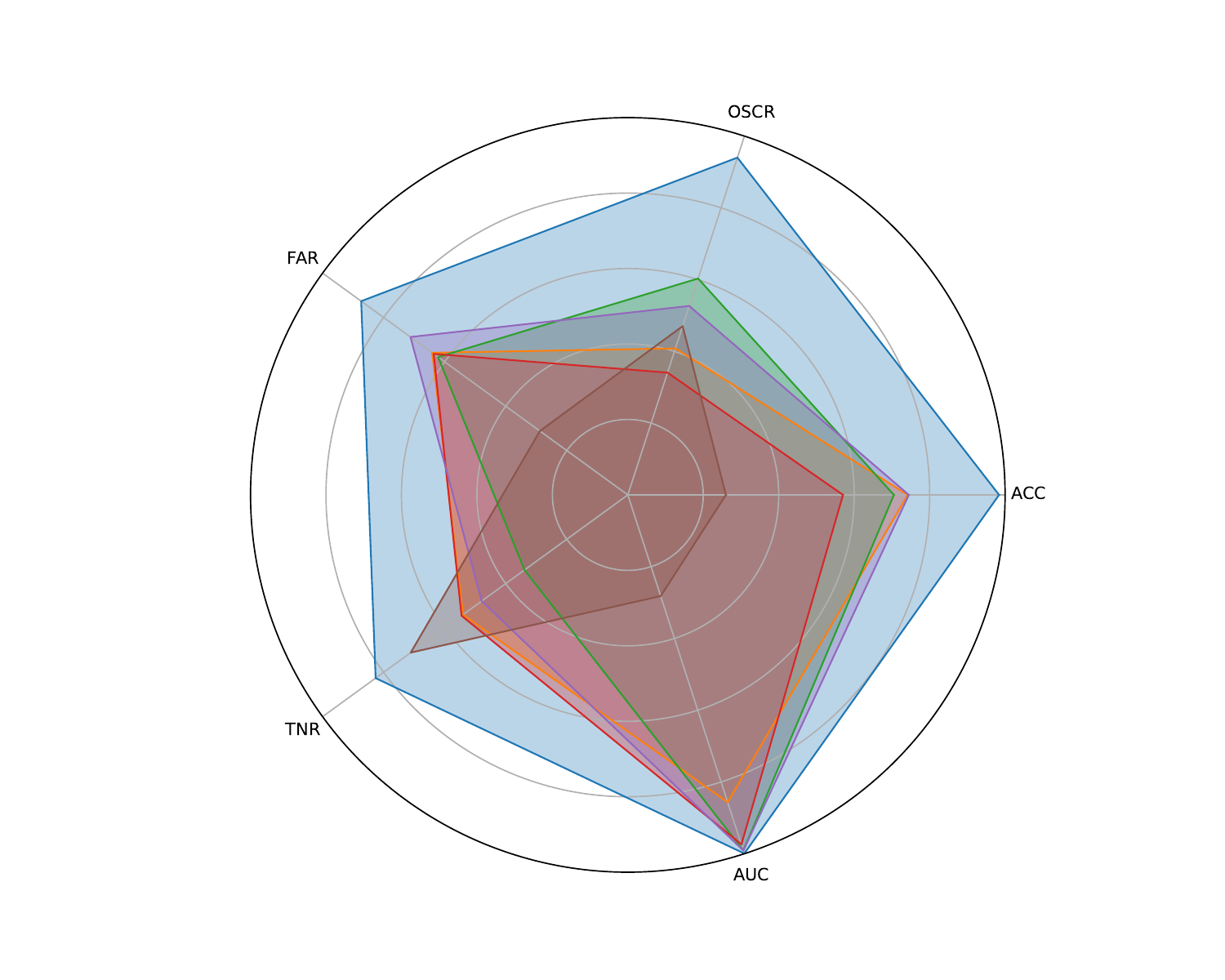}
        \caption{Experiment with MIT-BIH dataset}
        \label{fig:subfig2}
    \end{subfigure}
    \hfill
    \begin{subfigure}[b]{0.33\textwidth}
        \includegraphics[width=1.1\textwidth]{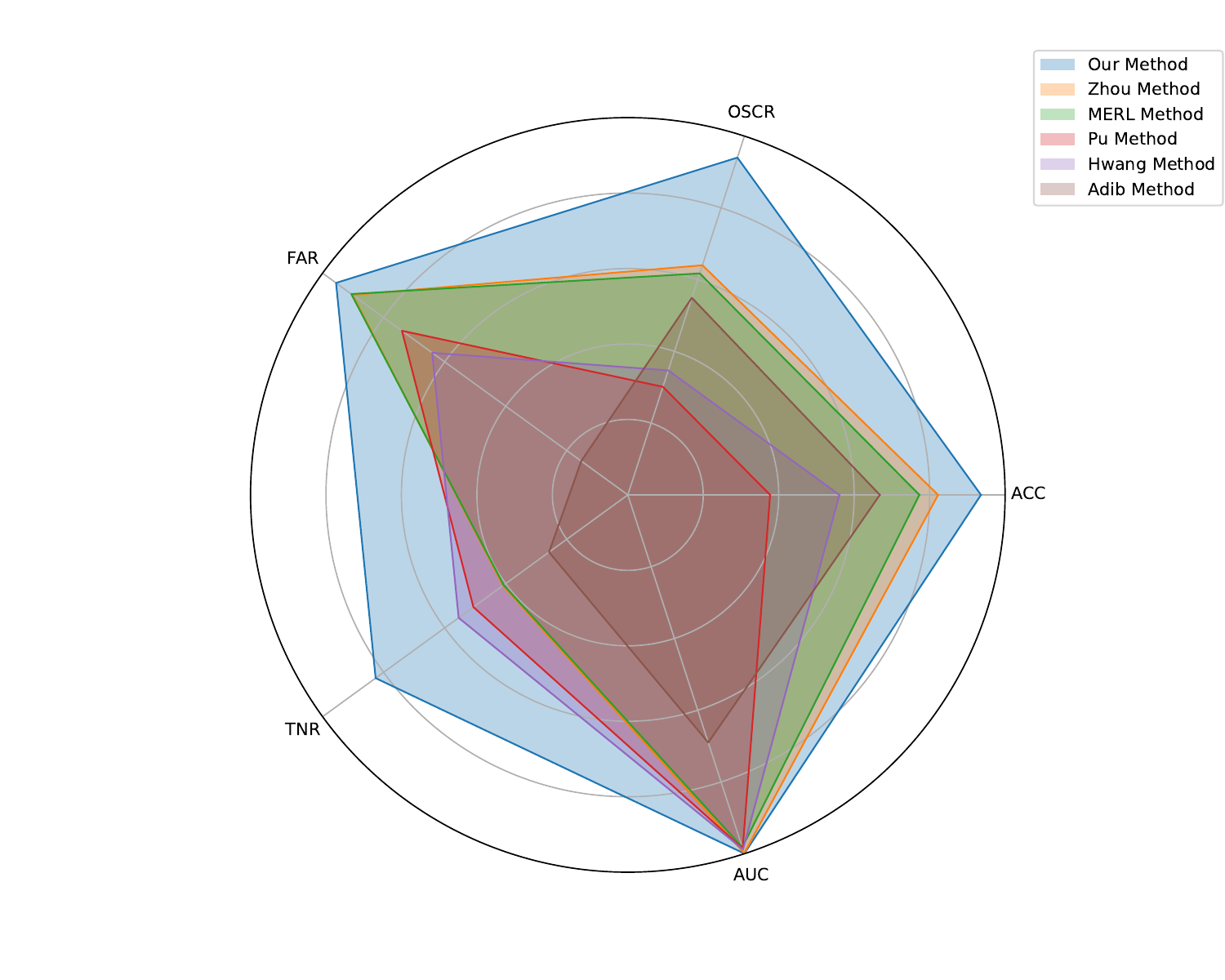}
        \caption{Experiment with Autonomic dataset}
        \label{fig:subfig3}
    \end{subfigure}
    \caption{The experimental results comparing various baseline methods are presented using ACC, OSCR, FAR, TNR, and AUC.}
    \label{fig:rader}
\end{figure*}

\textit{Dataset}. In the backbone comparative experiments, three distinct datasets were applied, each differing in the number of identity classes and sample sizes. Consequently, different data partitioning strategies were employed for each dataset. \textbf{ECGID} dataset, 41 available identity classes were selected, with 30 identity classes designated for enrolled users and 11 classes for the open set. Each sample was preprocessed by extracting 500 sampling points before and after the R-peak. \textbf{MIT-BIH} dataset, with 30 identity classes for enrolled users and 18 identity classes for the open set, maintaining the same preprocessing method as ECGID. \textbf{Autonomic} dataset, 100 identity classes were allocated for enrolled users and 50 identity classes for the open set, with 1000 sampling points extracted per sample. 

\textit{Encoder Structure}. In the experiments, different network architectures were used as the encoder for ECG signals, with representative models selected from ResNet and Vision Transformer (ViT). The ResNet-based encoders were configured with three different depths: [18, 34, 50], all employing 1D convolution (conv1D) as the primary convolutional operation. For ViT, conv1D was used to extract embedding features, and the full model consisted of 12 transformer blocks. The ViT-Tiny variant featured 3 attention heads, a multilayer perceptron (MLP) hidden dimension of 768, and an embedding dimension of 192. The ViT-Small variant had 6 attention heads, an MLP hidden dimension of 1536, and an embedding dimension of 384. The ViT-Base variant was designed with 12 attention heads, an MLP hidden dimension of 3072, and an embedding dimension of 768.

Table \ref{backbone} presents the comparative results of different backbone architectures for ECG signal encoder. Overall, models utilizing the ResNet backbone outperform those based on ViT, particularly in open-set evaluation metrics. In terms of network complexity, both backbone types exhibit performance degradation as model complexity increases, with ViT being more significantly affected. Notably, on the ECGID dataset, the ViT-Base backbone achieves an ACC of only 41.7\%. For the ResNet backbone, ResNet18 consistently achieves the best performance across most metrics on all datasets, indicating that convolutional networks are well-suited for capturing the waveform characteristics of ECG signals. In all subsequent experiments, the ResNet-based model is employed as the ECG signal encoder.

\subsection{Comparative Experiment}

\textit{Comparative Method}. To ensure a fair comparison, the experimental settings for all baseline methods were aligned with those of the proposed approach. All selected baseline methods were evaluated on ECG classification. The Adib method \cite{adib2022arrhythmia} employs a generative adversarial network to address class imbalance, and we adopted its classification model for comparison. The Pu method \cite{pu2023arrhythmia} utilizes a highly generalizable binary neural network for classification. The Zhou method \cite{zhou2024open} leverages hard negative samples and multi-hypersphere learning to improve the capability of ECG signal encoder. The MERL method \cite{liu2024zero} integrates multimodal contrastive learning between clinical text and ECG signals, as well as unimodal contrastive learning within the ECG modality, to enhance feature extraction capabilities. The Hwang method \cite{hwang2023multi} employs a ResNet-DenseNet architecture for multi-label classification tasks.

Figure \ref{fig:rader} presents a visualization of key performance metrics across different comparative methods under three datasets. Across evaluations on three benchmark datasets, our proposed method consistently outperforms all baseline approaches across all key performance metrics. Regardless of dataset size, our method consistently achieves stable ACC. On the ECGID dataset, our method attains a TNR of 49.03\% and a FAR as low as 5.39\%, indicating its strong capability to filter out most unregistered samples. While the Zhou method and MERL demonstrate higher ACC compared to other comparative methods due to their well-designed classifier tailored for identity authentication. However, these two methods exhibit nearly a 20\% gap in TNR compared to our method. This suggests that these methods struggle to accurately distinguish between registered and unregistered users. Our method not only maintains a high ACC for registered users but also effectively excludes the majority of unregistered users. By leveraging clear decision boundaries, our method ensures a strong balance between accurate identity verification and open-set sample rejection.

\begin{table*}[h!]
\caption{
Results of comparative experiments under varying proportions of open-set data. The performance is assessed using the OSCR and the FAR to reflect the model’s effectiveness in distinguishing between registered and unregistered identities.}
\label{bigtable}
\centering
\renewcommand{\arraystretch}{1.2} 
\setlength{\tabcolsep}{5pt} 
\small 
\begin{tabular}{c||cc|cc|cc|cc|cc}
\toprule
\multirow{3}{*}{Method} & \multicolumn{10}{c}{Proportion} \\  
 & \multicolumn{2}{c}{1:1} & \multicolumn{2}{c}{1:2} & \multicolumn{2}{c}{1:3} & \multicolumn{2}{c}{1:5} & \multicolumn{2}{c}{1:10} \\ \cline{2-11}
 & OSCR[\%] & FAR[\%] & OSCR[\%] & FAR[\%] & OSCR[\%] & FAR[\%] & OSCR[\%] & FAR[\%] & OSCR[\%] & FAR[\%] \\ 
\midrule
Zhou  & 87.19 & 15.00 & 83.37 & 25.57 & 83.61 & 30.85 & 82.54 & 36.93 & 82.54 & 42.66 \\
MERL  & 88.78 & 15.24 & 88.72 & 24.67 & 90.36 & 29.87 & 89.13 & 36.19 & 89.07 & 42.20 \\
Pu    & 74.31 & 18.61 & 75.60 & 27.42 & 75.88 & 32.41 & 75.79 & 38.00 & 74.43 & 43.38 \\ 
Hwang & 73.87 & 18.36 & 69.37 & 28.66 & 70.15 & 33.36 & 68.16 & 38.91 & 67.69 & 43.87 \\
Adib  & 11.57 & 23.81 & 11.10 & 32.89 & 10.87 & 37.39 & 10.95 & 41.58 & 10.80 & 45.50  \\
Ours  & 97.96 & 12.92 & 95.53 & 23.17 & 95.98 & 28.80 & 95.03 & 35.36 & 95.27 & 41.68 \\
\bottomrule
\end{tabular}
\label{table:proportion}
\end{table*}

\subsection{Longer Proportion of Open-set Data} 

\textit{Dataset}. Due to the limited number of samples in the ECGID and MIT-BIH datasets, the available open-set data is relatively scarce. To ensure a more comprehensive evaluation, we selected the Autonomic dataset for this experiment, as it provides a sufficient amount of data. Within this dataset, we selected 30 identity labels as the close-set data, while the open-set data was constructed by selecting [30, 60, 90, 120, 150, 180, 210, 240, 270, 300] identity categories as comparisons. These selections correspond to open-set to closed-set ratios ranging from 1:1 to 1:10, enabling a systematic analysis of the model’s performance across varying levels of open-set complexity.

\begin{figure}[h] 
\centering 
\includegraphics[width=0.5\textwidth]{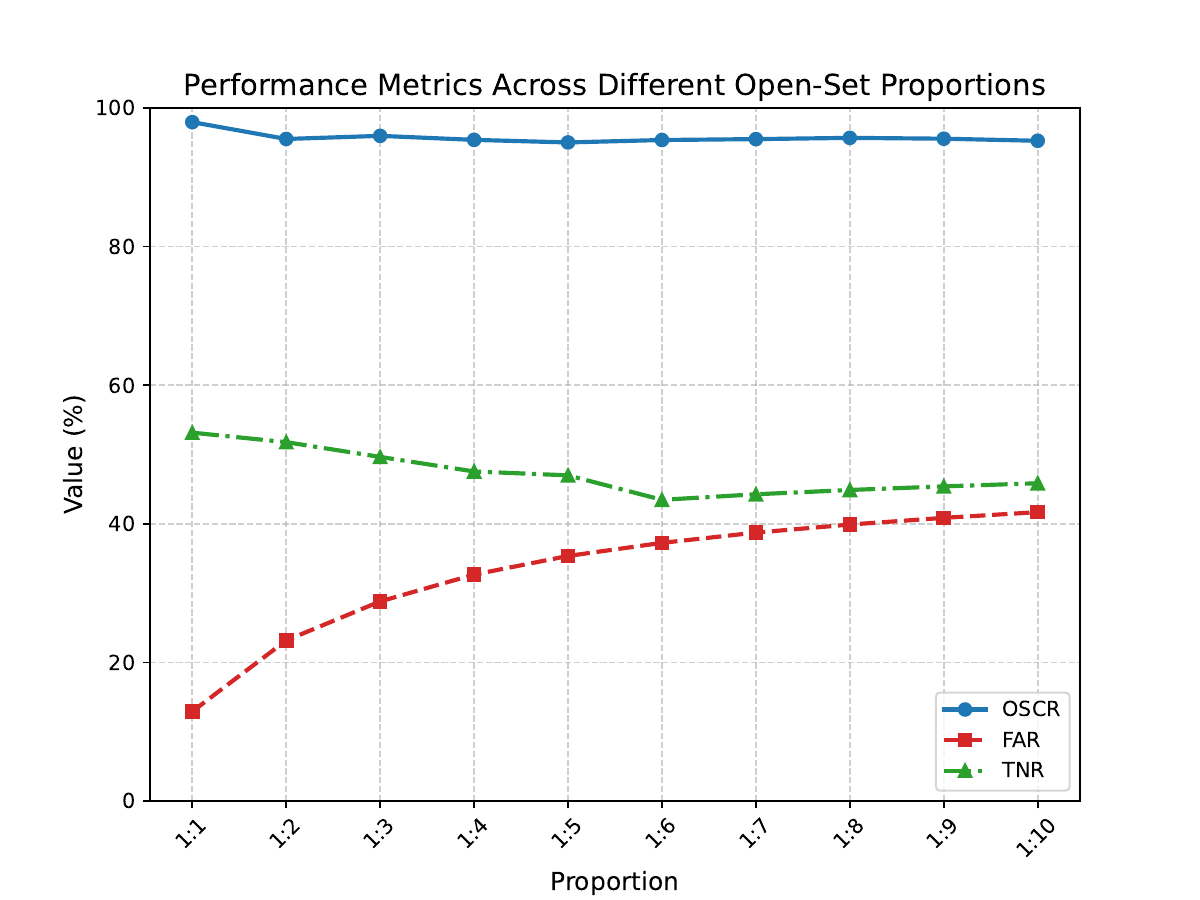} 
\caption{Line chart illustrating the variations in OSCR, FAR, and TNR as the ratio of open-set data to close-set data changes.} 
\label{fig:line} 
\end{figure}

To simulate the diversity of open-set data in real-world scenarios, we constructed open-set datasets with varying proportions to evaluate the model's ability to distinguish closed-set data in the presence of large-scale open-set samples. Figure \ref{fig:line} presents the results obtained under the aforementioned experimental settings. The model achieves an ACC of 99.83\% on the closed-set data, indicating its ability to correctly classify the vast majority of samples. Furthermore, across different open-set data proportions, the OSCR remains consistently above 95\%. This demonstrates the model's robustness in maintaining high recognition accuracy even in the presence of external data interference, highlighting its effectiveness in distinguishing between known and unknown classes. As the proportion of open-set data increases, the FAR exhibits a sharp rise before stabilizing around 40\%, while the TNR shows a declining trend, eventually settling at approximately 45\%. This indicates that when the open-set dataset remains within a reasonable range, our method effectively identifies registered users with high accuracy. However, if the open-set data surpasses a certain threshold, the model may become less effective at distinguishing unregistered users, potentially leading to an increased acceptance of unauthorized samples.

Table \ref{table:proportion} presents the experimental results comparing our method with multiple baseline approaches under varying open-set proportion settings. The evaluation metrics include OSCR and FAR, which are used to assess the model's ability to accurately recognize registered users in an open-set environment and to quantify the proportion of unregistered users mistakenly accepted. In the closed-set dataset, Zhou's method achieved an ACC of 94.56\%, MERL attained 97.00\%, Pu reached 92.18\%, Hwang obtained 94.14\%, and Adib achieved 24.00\%. Our proposed method achieved an ACC of 99.83\%, outperforming all other approaches and demonstrating superior recognition capability in a closed-set scenario. As the proportion of open-set data increases, all methods exhibit a rise in the FAR and a decline in the OSCR. However, across all experimental settings, our method consistently achieves the highest and most stable OSCR compared to all comparative methods. This indicates that our method effectively identifies registered users even in the presence of open-set data. Furthermore, in scenarios where open-set data is less prevalent, our method demonstrates the lowest FAR, successfully rejecting the majority of unseen samples. These results highlight the robustness and efficiency of our method in balancing open-set recognition and false acceptance mitigation, making it particularly suitable for real-world applications where reliable user authentication is critical.

\subsection{Ablation Study}

\begin{figure*}[htbp]
    \centering
    \begin{subfigure}[b]{0.24\textwidth}
        \includegraphics[width=1.15\textwidth]{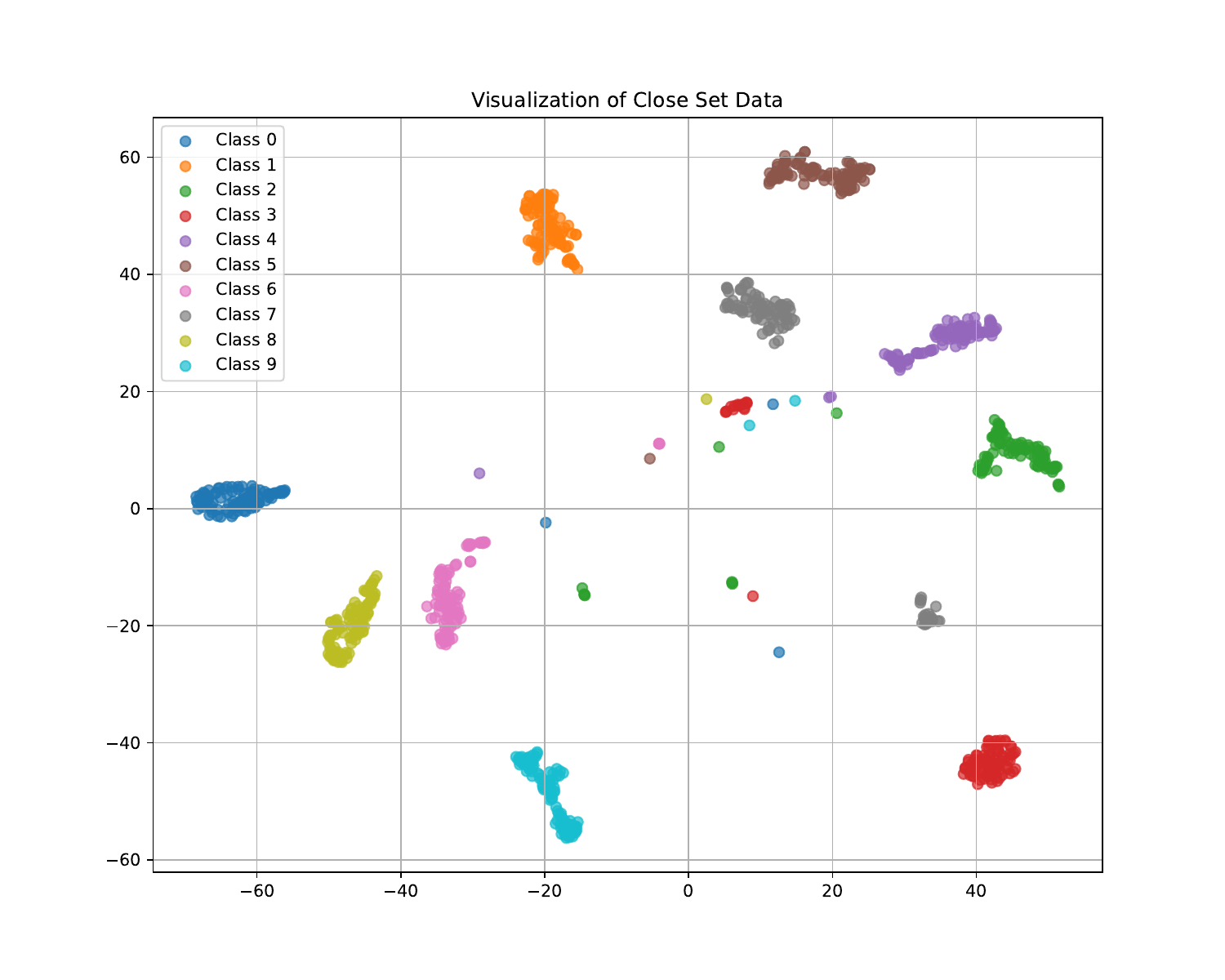}
        \caption{Close-set data in omplete model}
        \label{fig:subfig1}
    \end{subfigure}
    \hfill
    \begin{subfigure}[b]{0.24\textwidth}
        \includegraphics[width=1.15\textwidth]{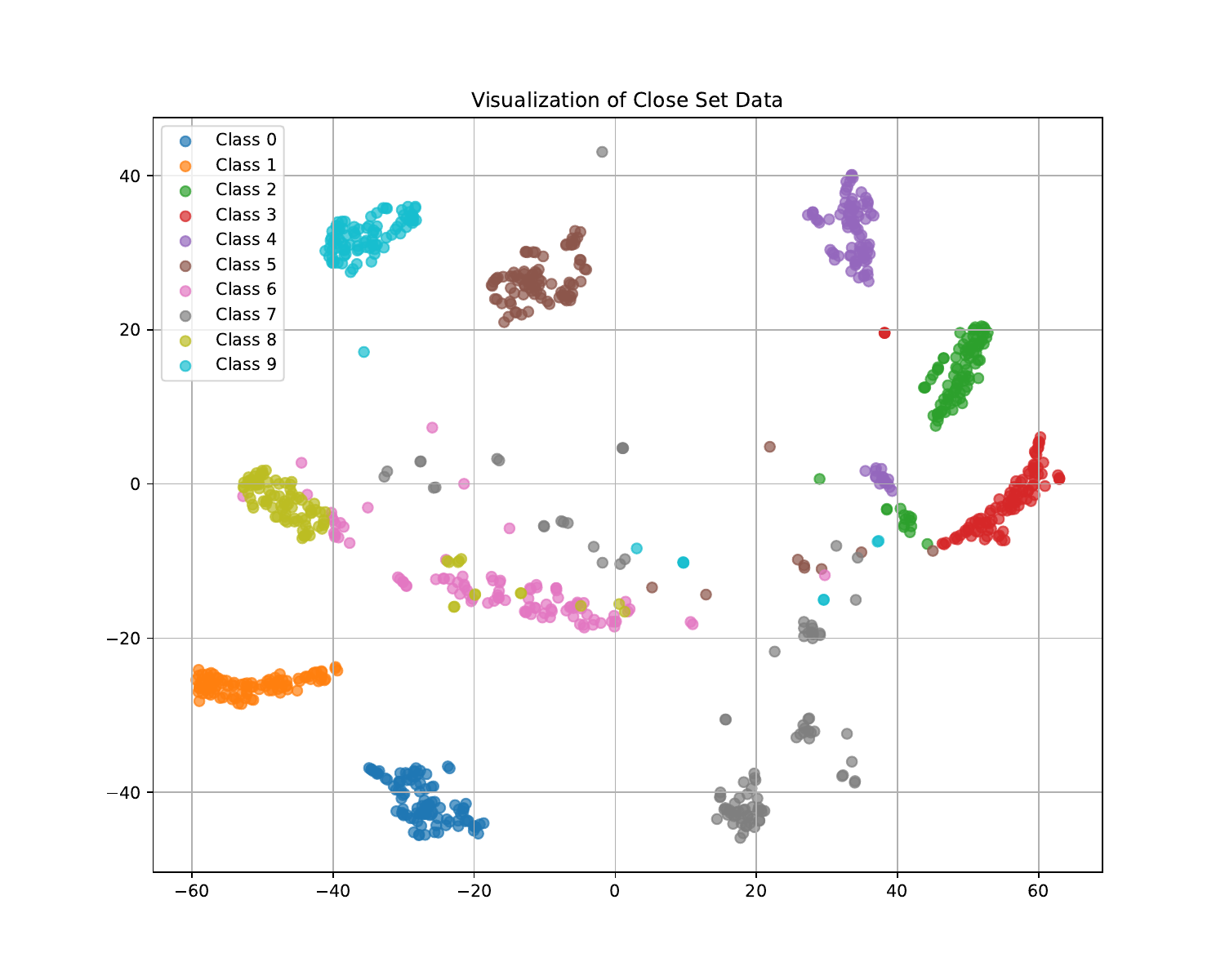}
        \caption{Close-set data in model only with B.1}
        \label{fig:subfig2}
    \end{subfigure}
    \hfill
    \begin{subfigure}[b]{0.24\textwidth}
        \includegraphics[width=1.15\textwidth]{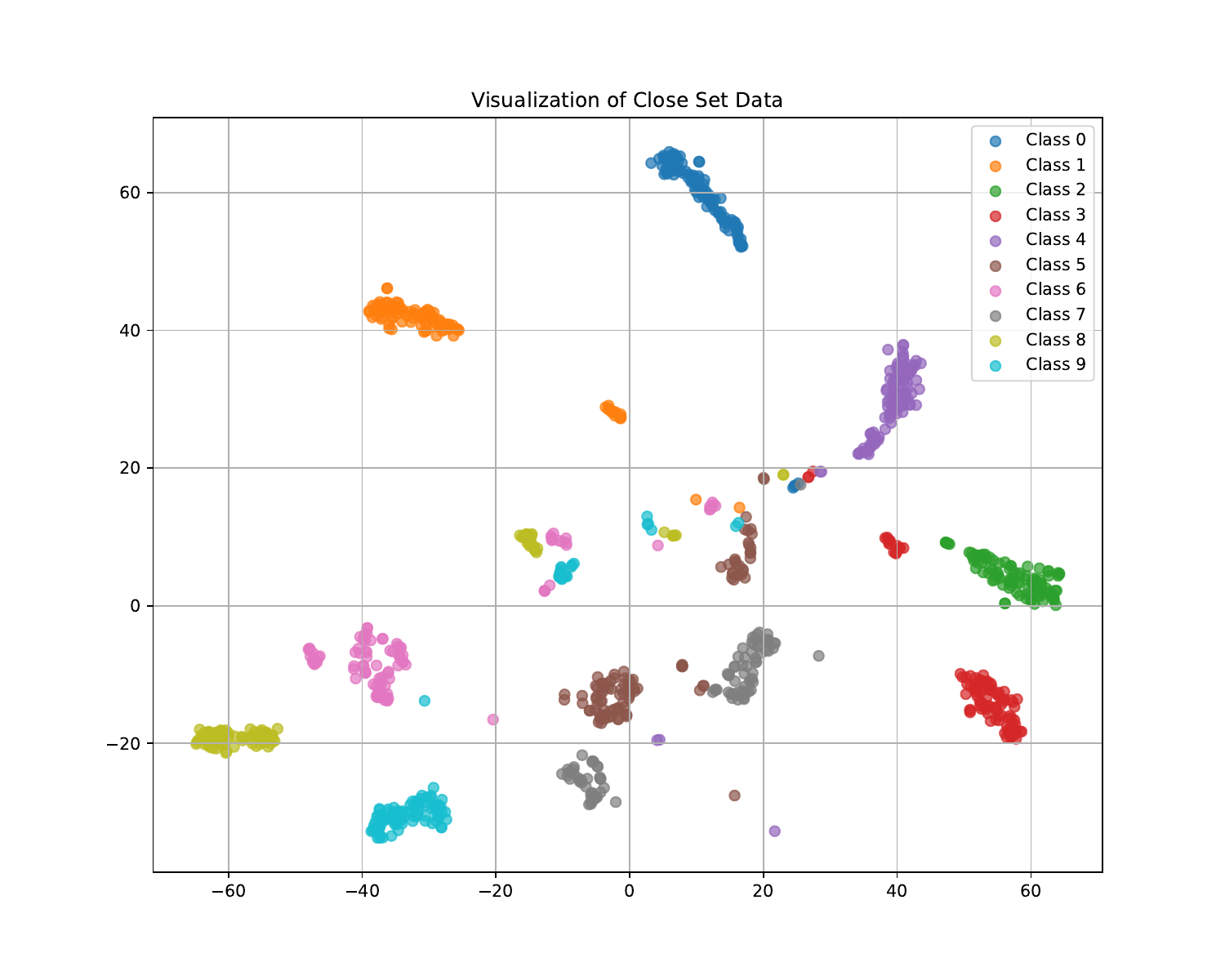}
        \caption{Close-set data in model only with B.2}
        \label{fig:subfig3}
    \end{subfigure}
    \hfill
    \begin{subfigure}[b]{0.24\textwidth}
        \includegraphics[width=1.15\textwidth]{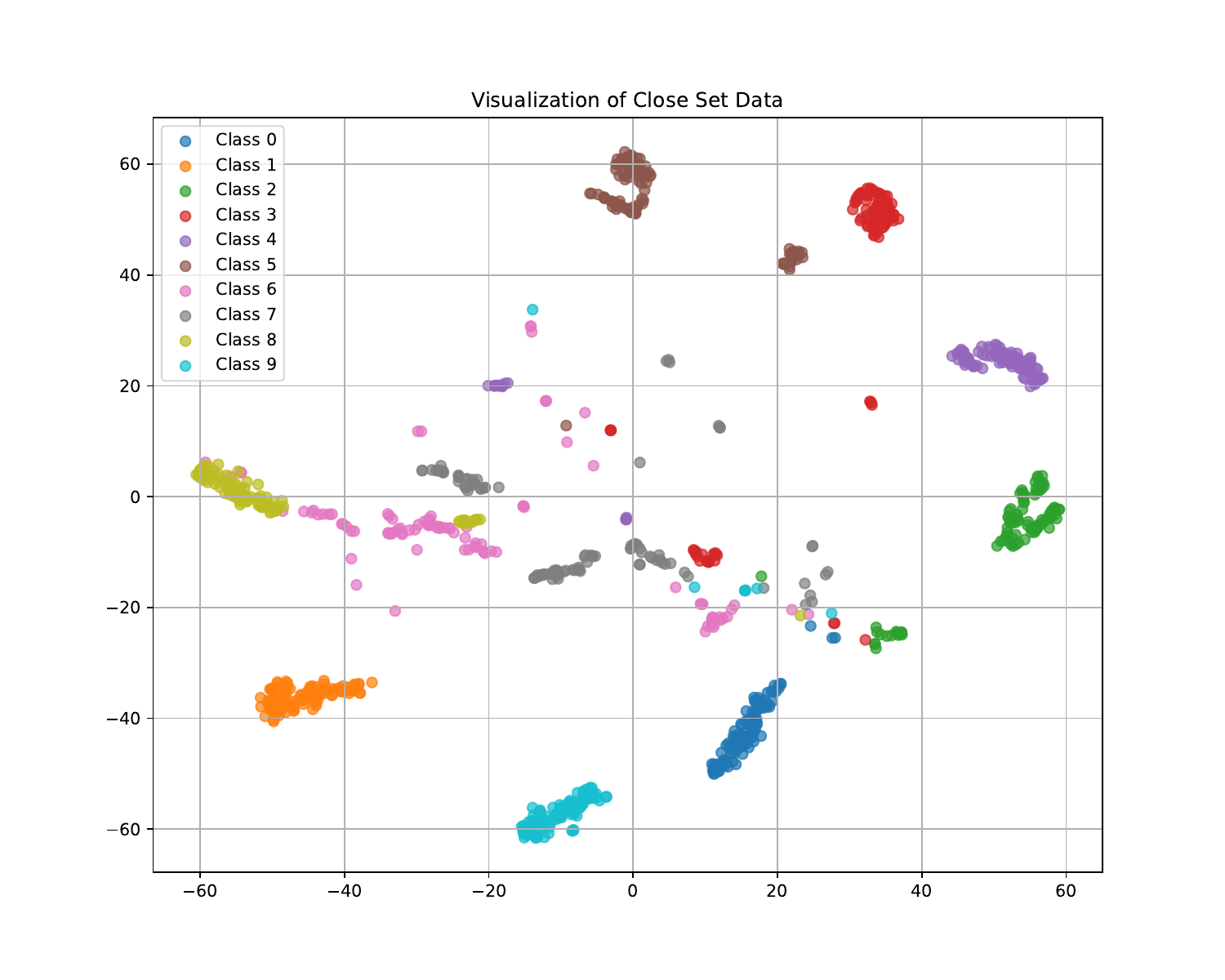}
        \caption{Close-set data in model only with B.3}
        \label{fig:subfig4}
    \end{subfigure}
    \\
    \begin{subfigure}[b]{0.24\textwidth}
        \includegraphics[width=1.15\textwidth]{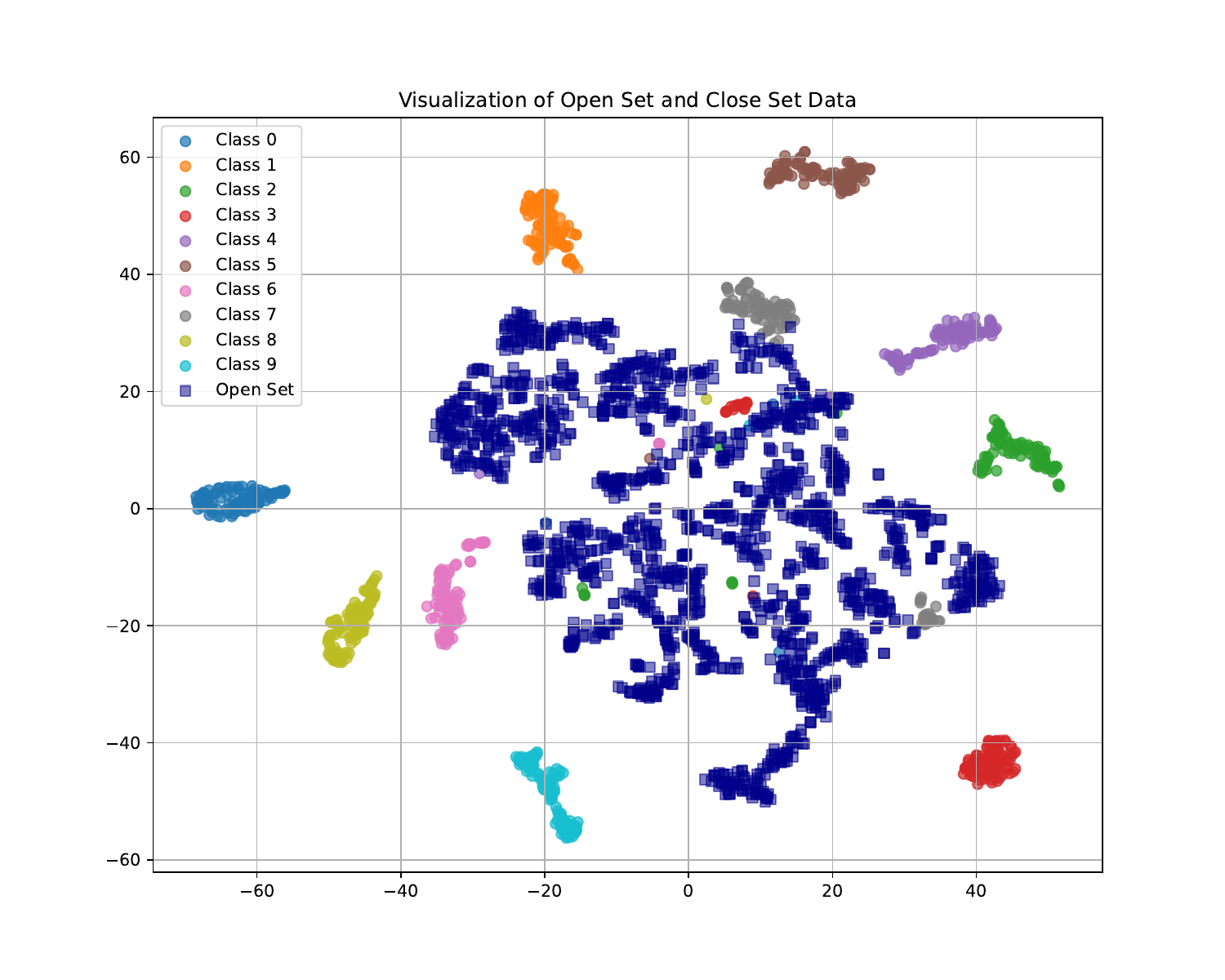}
        \caption{Open-set data in complete model}
        \label{fig:subfig5}
    \end{subfigure}
    \hfill
    \begin{subfigure}[b]{0.24\textwidth}
        \includegraphics[width=1.15\textwidth]{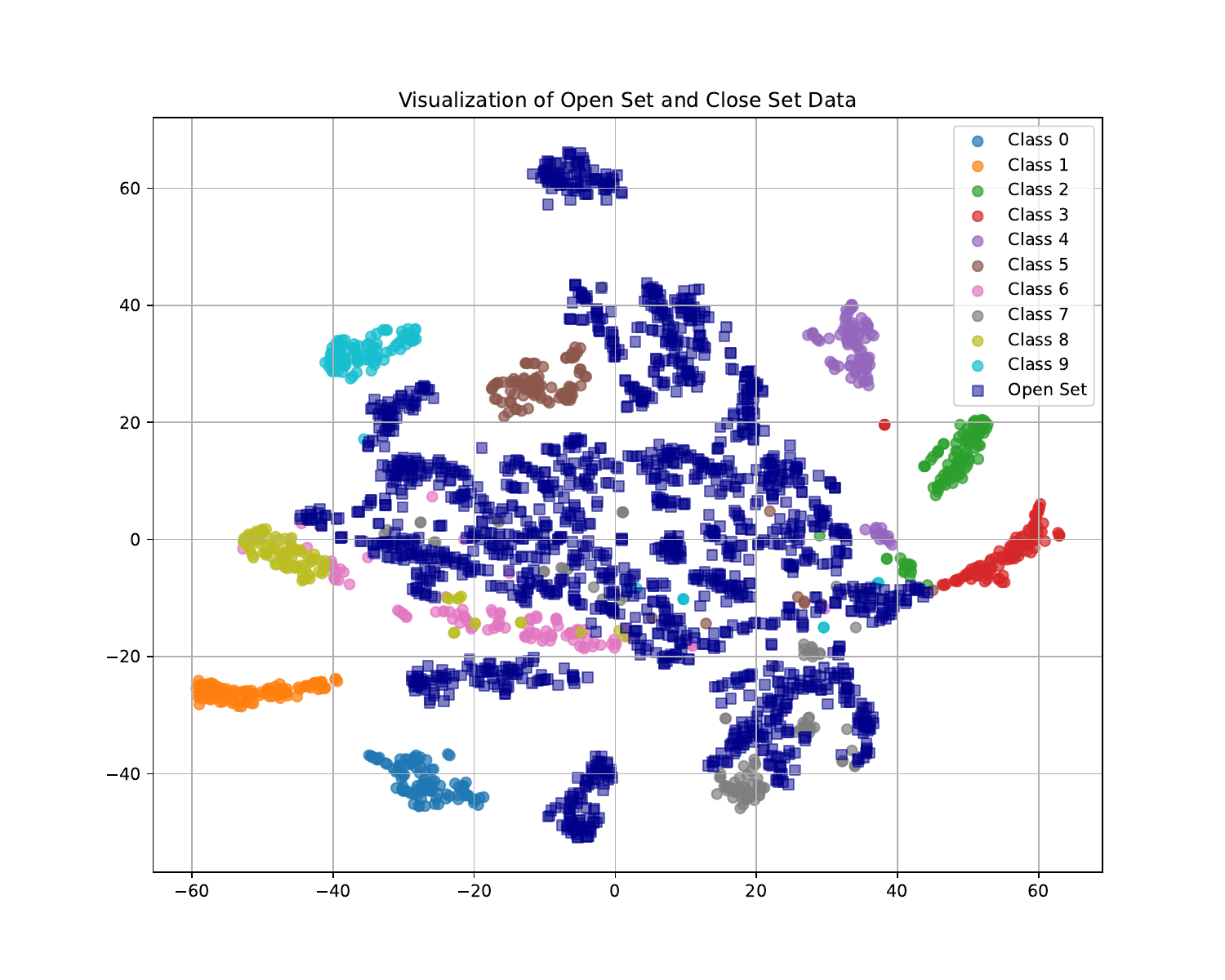}
        \caption{Open-set data in model only with B.1}
        \label{fig:subfig6}
    \end{subfigure}
    \hfill
    \begin{subfigure}[b]{0.24\textwidth}
        \includegraphics[width=1.15\textwidth]{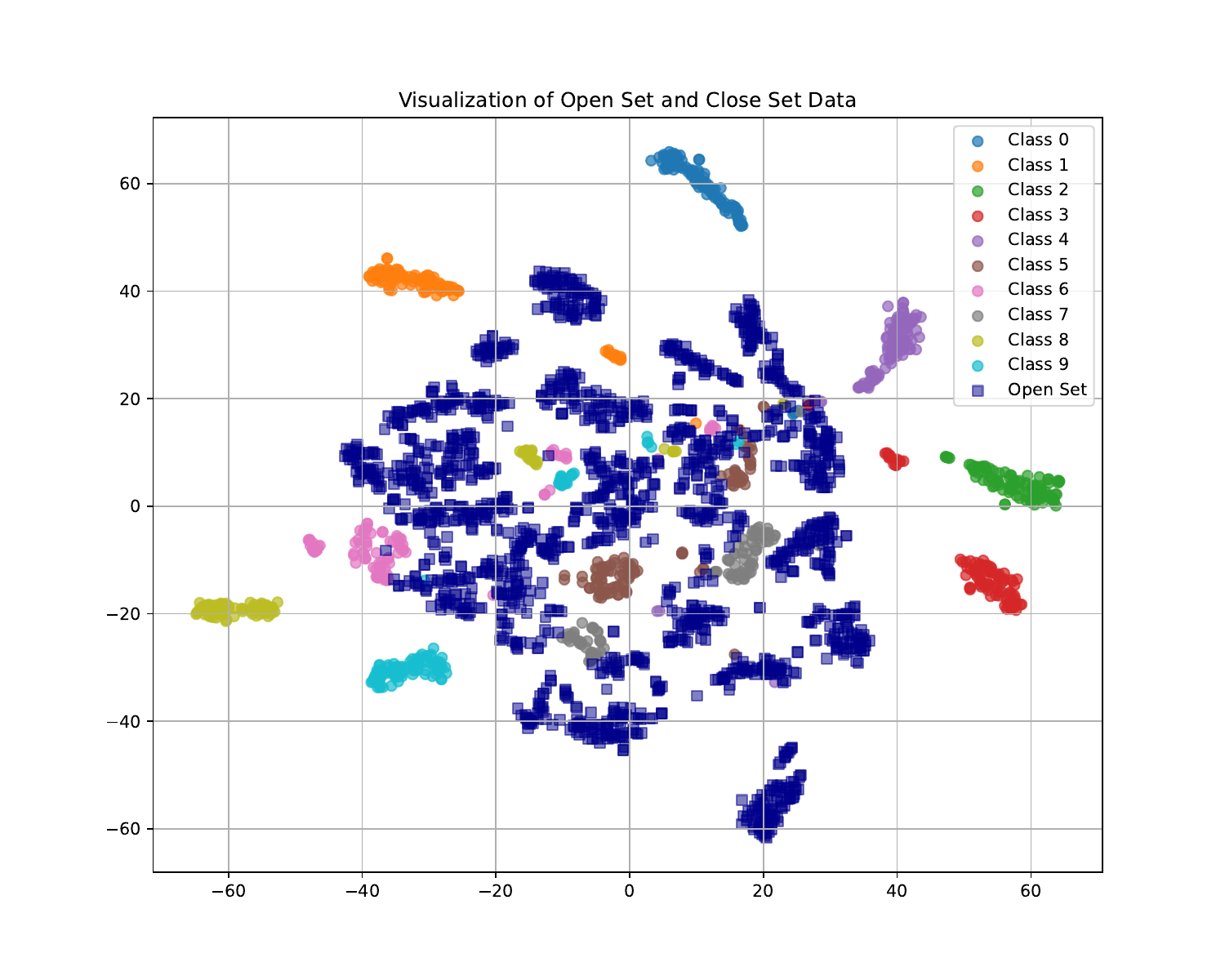}
        \caption{Open-set data in model only with B.2}
        \label{fig:subfig7}
    \end{subfigure}
    \hfill
    \begin{subfigure}[b]{0.24\textwidth}
        \includegraphics[width=1.15\textwidth]{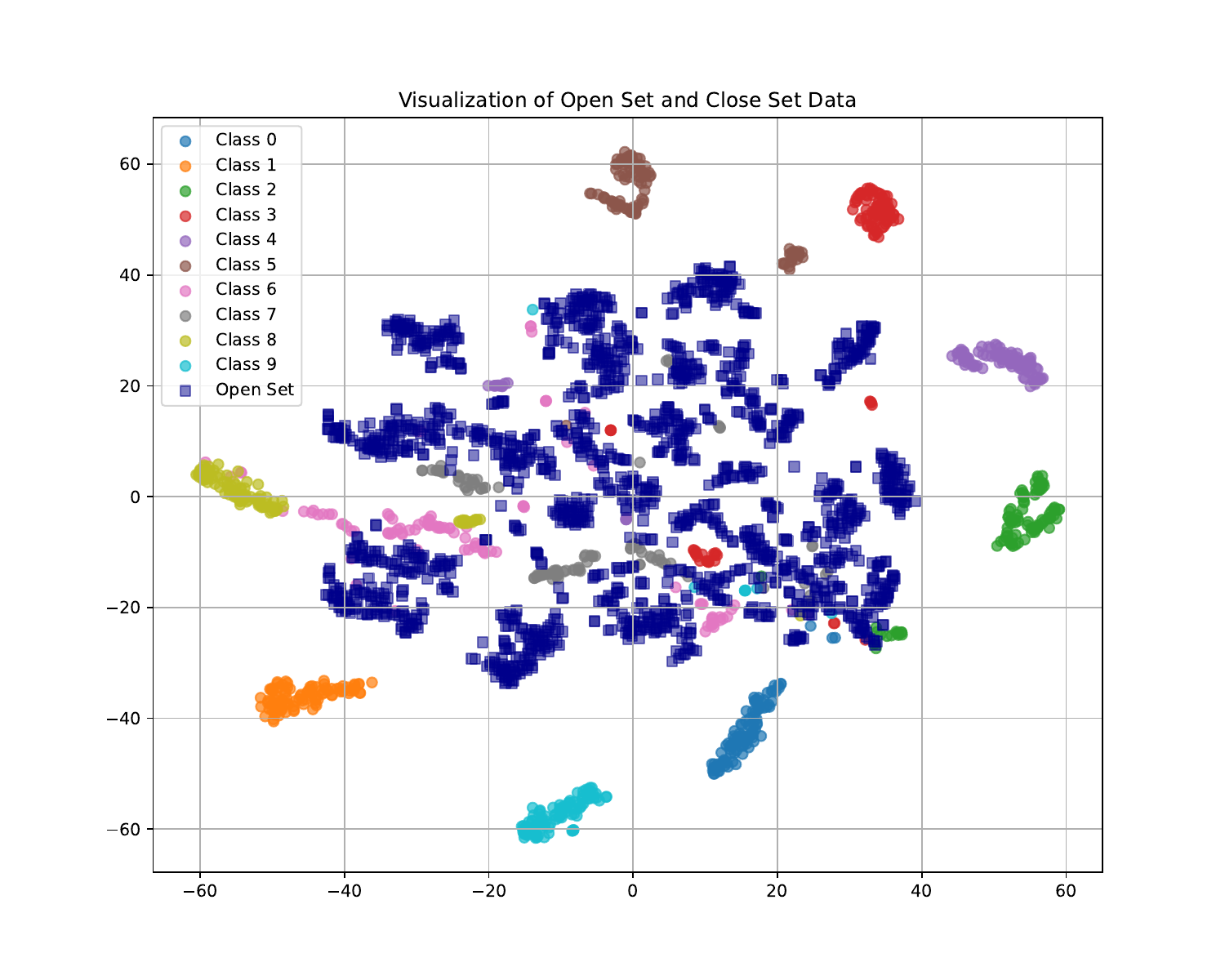}
        \caption{Open-set data in model only with B.3}
        \label{fig:subfig8}
    \end{subfigure}
    \caption{T-SNE visualizations of sample feature distributions under different ablation settings. Notably, (b) and (f) represent the results obtained using only the irrelevant sample repulsion learning module (only with B.1). (c) and (g) represent the results obtained using only the self-constraint center learning module (only with B.2). (d) and (h) represent the results obtained using only the dynamic prototype learning learning module (only with B.3).}
    \label{fig:tsne}
\end{figure*}

Table \ref{tab:ablation} presents the results of several ablation studies. The multimodal pretraining significantly enhances the model's recognition accuracy. Additionally, the irrelevant sample repulsion learning module and the self-constraint center learning module effectively reduce the probability of open-set samples being misclassified as registered users.

In the ablation studies of each component, the softmax of classification loss in our method was computed using probabilities derived from the distance between samples and their corresponding prototypes. In contrast, in the baseline experiments without dynamic prototype learning, the standard cross-entropy loss with softmax was used to replace the distance.

As shown in Table \ref{tab:ablation}, the model achieved nearly 18\% higher ACC in closed-set recognition after multi-modal pre-training compared to the model without pre-training. It also contributed to worse performance in open-set metrics. The results demonstrate that the signal encoder, after multimodal pretraining, is able to capture identity-related information more effectively, thereby providing a stronger foundation for downstream identity authentication tasks.

Self-constraint center learning is a crucial module that effectively brings intra-class samples closer to the center of the corresponding identity label while reducing the dispersion of sample distributions. As a result, models without self-constraint center learning exhibit higher FAR, leading to the misclassification of some open-set samples as registered users. The effect of dynamic prototypes is similar to that of self-constraint center learning; however, the best results are achieved when both are applied simultaneously. The presence of irrelevant sample repulsion learning helps to push the registered samples further away from the dispersed open-set samples, and without irrelevant sample repulsion learning, FAR tends to increase slightly. The model achieves optimal performance only when all three sample distribution constraint methods are employed together.

Figure \ref{fig:tsne} illustrates the dimensionality-reduced feature distributions in the ablation study. To facilitate observation of the sample distributions, subfigures (\ref{fig:subfig1})–(\ref{fig:subfig4}) depict the distributions of samples within the closed set, while subfigures (\ref{fig:subfig5})–(\ref{fig:subfig8}) show the distributions of all samples when open-set data is included.

Subfigure (\ref{fig:subfig1}) and (\ref{fig:subfig5}) present the distributions using the complete proposed method, where the open-set data is clearly confined to a smaller area, with other close-set samples remaining relatively distant from the open-set sample distribution. In contrast, subfigures (\ref{fig:subfig2}) and (\ref{fig:subfig6}) show the outcome when only irrelevant sample repulsion learning and cross-entropy loss are used to guide model convergence. This approach also achieves a high ACC, successfully extracting the feature distribution of unknown labels, but results in a more scattered open-set sample distribution. A similar effect is observed when only the dynamic prototype method is applied, as shown in subfigures (\ref{fig:subfig4}) and (\ref{fig:subfig8}). However, when only the self-constraint method and cross-entropy loss are used, as depicted in subfigures (\ref{fig:subfig3}) and (\ref{fig:subfig7}), the model performs poorly in classification and produces a highly dispersed open-set samples distribution.


\begin{table}[h!]
\centering
\small 
\caption{Ablation study of the method. A denotes whether multi-modal pre-training is applied. B.1 denotes whether the reciprocal point is applied. B.2 denotes whether dynamic prototype is applied. B.3 denotes whether self-constraint center learning is applied.}
\label{tab:ablation}
\begin{tabular}{c||c|c|c||c|c|c}
\toprule
A & B.1 & B.2 & B.3 & ACC[\%] & OSCR[\%] & FAR[\%] \\
\midrule
$\times$ & $\checkmark$ & $\checkmark$ & $\checkmark$ & 80.10 & 64.02 & 19.44  \\
\midrule
$\checkmark$ & $\checkmark$ & $\checkmark$ & $\checkmark$ & 99.60 & 97.60 & \textbf{7.53} \\
$\checkmark$ & $\checkmark$ & $\times$ & $\times$ & 99.53 & 97.16 & 15.58  \\
$\checkmark$ & $\times$ & $\checkmark$ & $\times$ & 99.63 & 97.57 & 7.56  \\
$\checkmark$ & $\times$ & $\times$ & $\checkmark$ & 99.57 & 96.56 & 7.78  \\
$\checkmark$ & $\checkmark$ & $\checkmark$ & $\times$ & 99.53 & 97.12 & 15.11  \\
$\checkmark$ & $\checkmark$ & $\times$ & $\checkmark$ & 99.70 & 97.51 & 15.48  \\
$\checkmark$ & $\times$ & $\checkmark$ & $\checkmark$ & 99.13 & 93.68 & 8.40  \\
\bottomrule
\end{tabular}
\end{table}


\section{Summary and conclusions}
This paper presents a novel ECG identity authentication method designed for open-set scenarios. The proposed method has been rigorously evaluated under varying proportions of open-set data. A key innovation of our method is the incorporation of contrastive learning with multi-modal data during pretraining, where ECG signals and text reports based on the fiducial feature are integrated to enhance the signal encoder’s ability to represent ECG features comprehensively.

During the fine-tuning phase for the downstream identity authentication task, we introduce Self-constraint Center Learning, which further compacts the feature representations into a more discriminative subspace, leading to an identity recognition accuracy of 99.83\%, surpassing comparative ECG classification methods. Additionally, we propose Irrelevant Sample Repulsion Learning, which effectively restricts the distribution of unseen open-set samples to a more constrained space, enabling the model to efficiently filter out unregistered identities, achieving a FAR as low as 5.39\%. 

Extensive experimental results demonstrate that our method maintains highly effective identity authentication performance even in the presence of large-scale open-set data, establishing a new benchmark for ECG-based authentication in open-world settings. The ablation studies confirm the effectiveness of the proposed modules in enhancing identity recognition under open-set conditions. Incorporating all modules leads to a significant reduction in the FAR and a notable improvement in the OSCR.

However, current research still exhibits certain limitations. Specifically, when a large volume of open-set data is present, existing models struggle to effectively reject the majority of unregistered users. Consequently, future research efforts will focus on developing strategies to further reduce the FAR.





\bibliographystyle{elsarticle-harv} 
\bibliography{main}

\begin{thebibliography}{34}
\expandafter\ifx\csname natexlab\endcsname\relax\def\natexlab#1{#1}\fi
\providecommand{\url}[1]{\texttt{#1}}
\providecommand{\href}[2]{#2}
\providecommand{\path}[1]{#1}
\providecommand{\DOIprefix}{doi:}
\providecommand{\ArXivprefix}{arXiv:}
\providecommand{\URLprefix}{URL: }
\providecommand{\Pubmedprefix}{pmid:}
\providecommand{\doi}[1]{\href{http://dx.doi.org/#1}{\path{#1}}}
\providecommand{\Pubmed}[1]{\href{pmid:#1}{\path{#1}}}
\providecommand{\bibinfo}[2]{#2}
\ifx\xfnm\relax \def\xfnm[#1]{\unskip,\space#1}\fi
\bibitem[{Adib et~al.(2022)Adib, Afghah and Prevost}]{adib2022arrhythmia}
\bibinfo{author}{Adib, E.}, \bibinfo{author}{Afghah, F.}, \bibinfo{author}{Prevost, J.J.}, \bibinfo{year}{2022}.
\newblock \bibinfo{title}{Arrhythmia classification using cgan-augmented ecg signals}, in: \bibinfo{booktitle}{2022 IEEE International Conference on Bioinformatics and Biomedicine (BIBM)}, \bibinfo{organization}{IEEE}. pp. \bibinfo{pages}{1865--1872}.
\bibitem[{Aslan and Choi(2024)}]{aslan2024visgin}
\bibinfo{author}{Aslan, H.{\.I}.}, \bibinfo{author}{Choi, C.}, \bibinfo{year}{2024}.
\newblock \bibinfo{title}{Visgin: Visibility graph neural network on one-dimensional data for biometric authentication}.
\newblock \bibinfo{journal}{Expert Systems with Applications} \bibinfo{volume}{237}, \bibinfo{pages}{121323}.
\bibitem[{Boumbarov et~al.(2009)Boumbarov, Velchev and Sokolov}]{boumbarov2009ecg}
\bibinfo{author}{Boumbarov, O.}, \bibinfo{author}{Velchev, Y.}, \bibinfo{author}{Sokolov, S.}, \bibinfo{year}{2009}.
\newblock \bibinfo{title}{Ecg personal identification in subspaces using radial basis neural networks}, in: \bibinfo{booktitle}{2009 IEEE international workshop on intelligent data acquisition and advanced computing systems: technology and applications}, \bibinfo{organization}{IEEE}. pp. \bibinfo{pages}{446--451}.
\bibitem[{Chan et~al.(2008)Chan, Hamdy, Badre and Badee}]{chan2008wavelet}
\bibinfo{author}{Chan, A.D.}, \bibinfo{author}{Hamdy, M.M.}, \bibinfo{author}{Badre, A.}, \bibinfo{author}{Badee, V.}, \bibinfo{year}{2008}.
\newblock \bibinfo{title}{Wavelet distance measure for person identification using electrocardiograms}.
\newblock \bibinfo{journal}{IEEE transactions on instrumentation and measurement} \bibinfo{volume}{57}, \bibinfo{pages}{248--253}.
\bibitem[{Chen et~al.(2021)Chen, Peng, Wang and Tian}]{chen2021adversarial}
\bibinfo{author}{Chen, G.}, \bibinfo{author}{Peng, P.}, \bibinfo{author}{Wang, X.}, \bibinfo{author}{Tian, Y.}, \bibinfo{year}{2021}.
\newblock \bibinfo{title}{Adversarial reciprocal points learning for open set recognition}.
\newblock \bibinfo{journal}{IEEE Transactions on Pattern Analysis and Machine Intelligence} \bibinfo{volume}{44}, \bibinfo{pages}{8065--8081}.
\bibitem[{Dhamija et~al.(2018)Dhamija, G{\"u}nther and Boult}]{dhamija2018reducing}
\bibinfo{author}{Dhamija, A.R.}, \bibinfo{author}{G{\"u}nther, M.}, \bibinfo{author}{Boult, T.}, \bibinfo{year}{2018}.
\newblock \bibinfo{title}{Reducing network agnostophobia}.
\newblock \bibinfo{journal}{Advances in Neural Information Processing Systems} \bibinfo{volume}{31}.
\bibitem[{Gow et~al.(2023)Gow, Pollard, Nathanson, Johnson, Moody, Fernandes, Greenbaum, Waks, Eslami, Carbonati et~al.}]{gow2023mimic}
\bibinfo{author}{Gow, B.}, \bibinfo{author}{Pollard, T.}, \bibinfo{author}{Nathanson, L.A.}, \bibinfo{author}{Johnson, A.}, \bibinfo{author}{Moody, B.}, \bibinfo{author}{Fernandes, C.}, \bibinfo{author}{Greenbaum, N.}, \bibinfo{author}{Waks, J.W.}, \bibinfo{author}{Eslami, P.}, \bibinfo{author}{Carbonati, T.}, et~al., \bibinfo{year}{2023}.
\newblock \bibinfo{title}{Mimic-iv-ecg: Diagnostic electrocardiogram matched subset}.
\newblock \bibinfo{journal}{Type: dataset} \bibinfo{volume}{6}, \bibinfo{pages}{13--14}.
\bibitem[{Han et~al.(2022)Han, Wang, Chen, Chen, Guo, Liu, Tang, Xiao, Xu, Xu et~al.}]{han2022survey}
\bibinfo{author}{Han, K.}, \bibinfo{author}{Wang, Y.}, \bibinfo{author}{Chen, H.}, \bibinfo{author}{Chen, X.}, \bibinfo{author}{Guo, J.}, \bibinfo{author}{Liu, Z.}, \bibinfo{author}{Tang, Y.}, \bibinfo{author}{Xiao, A.}, \bibinfo{author}{Xu, C.}, \bibinfo{author}{Xu, Y.}, et~al., \bibinfo{year}{2022}.
\newblock \bibinfo{title}{A survey on vision transformer}.
\newblock \bibinfo{journal}{IEEE transactions on pattern analysis and machine intelligence} \bibinfo{volume}{45}, \bibinfo{pages}{87--110}.
\bibitem[{Hoekema et~al.(2001)Hoekema, Uijen and Van~Oosterom}]{hoekema2001geometrical}
\bibinfo{author}{Hoekema, R.}, \bibinfo{author}{Uijen, G.J.}, \bibinfo{author}{Van~Oosterom, A.}, \bibinfo{year}{2001}.
\newblock \bibinfo{title}{Geometrical aspects of the interindividual variability of multilead ecg recordings}.
\newblock \bibinfo{journal}{IEEE Transactions on Biomedical Engineering} \bibinfo{volume}{48}, \bibinfo{pages}{551--559}.
\bibitem[{Hwang et~al.(2023)Hwang, Cha, Heo, Cho and Park}]{hwang2023multi}
\bibinfo{author}{Hwang, S.}, \bibinfo{author}{Cha, J.}, \bibinfo{author}{Heo, J.}, \bibinfo{author}{Cho, S.}, \bibinfo{author}{Park, Y.}, \bibinfo{year}{2023}.
\newblock \bibinfo{title}{Multi-label ecg abnormality classification using a combined resnet-densenet architecture with resu blocks}, in: \bibinfo{booktitle}{2023 IEEE EMBS Special Topic Conference on Data Science and Engineering in Healthcare, Medicine and Biology}, \bibinfo{organization}{IEEE}. pp. \bibinfo{pages}{111--112}.
\bibitem[{Irvine et~al.(2008)Irvine, Israel, Scruggs and Worek}]{irvine2008eigenpulse}
\bibinfo{author}{Irvine, J.M.}, \bibinfo{author}{Israel, S.A.}, \bibinfo{author}{Scruggs, W.T.}, \bibinfo{author}{Worek, W.J.}, \bibinfo{year}{2008}.
\newblock \bibinfo{title}{eigenpulse: Robust human identification from cardiovascular function}.
\newblock \bibinfo{journal}{Pattern Recognition} \bibinfo{volume}{41}, \bibinfo{pages}{3427--3435}.
\bibitem[{Jin et~al.(2025)Jin, Wang, Li, Li, Pan and Hong}]{jin2025reading}
\bibinfo{author}{Jin, J.}, \bibinfo{author}{Wang, H.}, \bibinfo{author}{Li, H.}, \bibinfo{author}{Li, J.}, \bibinfo{author}{Pan, J.}, \bibinfo{author}{Hong, S.}, \bibinfo{year}{2025}.
\newblock \bibinfo{title}{Reading your heart: Learning ecg words and sentences via pre-training ecg language model}.
\newblock \bibinfo{journal}{arXiv preprint arXiv:2502.10707} .
\bibitem[{Jin et~al.(2023)Jin, Kim, Chen, Comeau, Yeganova, Wilbur and Lu}]{jin2023medcpt}
\bibinfo{author}{Jin, Q.}, \bibinfo{author}{Kim, W.}, \bibinfo{author}{Chen, Q.}, \bibinfo{author}{Comeau, D.C.}, \bibinfo{author}{Yeganova, L.}, \bibinfo{author}{Wilbur, W.J.}, \bibinfo{author}{Lu, Z.}, \bibinfo{year}{2023}.
\newblock \bibinfo{title}{Medcpt: Contrastive pre-trained transformers with large-scale pubmed search logs for zero-shot biomedical information retrieval}.
\newblock \bibinfo{journal}{Bioinformatics} \bibinfo{volume}{39}, \bibinfo{pages}{btad651}.
\bibitem[{Kinga et~al.(2015)Kinga, Adam et~al.}]{kinga2015method}
\bibinfo{author}{Kinga, D.}, \bibinfo{author}{Adam, J.B.}, et~al., \bibinfo{year}{2015}.
\newblock \bibinfo{title}{A method for stochastic optimization}, in: \bibinfo{booktitle}{International conference on learning representations (ICLR)}, \bibinfo{organization}{San Diego, California;}.
\bibitem[{Koonce(2021)}]{koonce2021resnet}
\bibinfo{author}{Koonce, B.}, \bibinfo{year}{2021}.
\newblock \bibinfo{title}{Resnet 50}, in: \bibinfo{booktitle}{Convolutional neural networks with swift for tensorflow: image recognition and dataset categorization}. \bibinfo{publisher}{Springer}, pp. \bibinfo{pages}{63--72}.
\bibitem[{Krishnamoorthy and Raju(2024)}]{krishnamoorthy2024deep}
\bibinfo{author}{Krishnamoorthy, L.}, \bibinfo{author}{Raju, A.S.}, \bibinfo{year}{2024}.
\newblock \bibinfo{title}{Deep ensemble of vgg, resnet and inception for multimodal authentication system}, in: \bibinfo{booktitle}{2024 Second International Conference on Networks, Multimedia and Information Technology (NMITCON)}, \bibinfo{organization}{IEEE}. pp. \bibinfo{pages}{1--6}.
\bibitem[{Lee et~al.(2018)Lee, Jeong, Park, Yun and Park}]{lee2018efficient}
\bibinfo{author}{Lee, S.}, \bibinfo{author}{Jeong, Y.}, \bibinfo{author}{Park, D.}, \bibinfo{author}{Yun, B.J.}, \bibinfo{author}{Park, K.H.}, \bibinfo{year}{2018}.
\newblock \bibinfo{title}{Efficient fiducial point detection of ecg qrs complex based on polygonal approximation}.
\newblock \bibinfo{journal}{Sensors} \bibinfo{volume}{18}, \bibinfo{pages}{4502}.
\bibitem[{Liu et~al.(2024)Liu, Wan, Ouyang, Shah, Bai and Arcucci}]{liu2024zero}
\bibinfo{author}{Liu, C.}, \bibinfo{author}{Wan, Z.}, \bibinfo{author}{Ouyang, C.}, \bibinfo{author}{Shah, A.}, \bibinfo{author}{Bai, W.}, \bibinfo{author}{Arcucci, R.}, \bibinfo{year}{2024}.
\newblock \bibinfo{title}{Zero-shot ecg classification with multimodal learning and test-time clinical knowledge enhancement}.
\newblock \bibinfo{journal}{arXiv preprint arXiv:2403.06659} .
\bibitem[{Lugovaya(2005)}]{lugovaya2005biometric}
\bibinfo{author}{Lugovaya, T.S.}, \bibinfo{year}{2005}.
\newblock \bibinfo{title}{Biometric human identification based on electrocardiogram}.
\newblock \bibinfo{journal}{Master's thesis, Faculty of Computing Technologies and Informatics, Electrotechnical University ‘LETI’, Saint-Petersburg, Russian Federation} .
\bibitem[{Moody and Mark(2001)}]{moody2001impact}
\bibinfo{author}{Moody, G.B.}, \bibinfo{author}{Mark, R.G.}, \bibinfo{year}{2001}.
\newblock \bibinfo{title}{The impact of the mit-bih arrhythmia database}.
\newblock \bibinfo{journal}{IEEE engineering in medicine and biology magazine} \bibinfo{volume}{20}, \bibinfo{pages}{45--50}.
\bibitem[{Pereira et~al.(2023)Pereira, Concei{\c{c}}{\~a}o, Sencadas and Sebasti{\~a}o}]{pereira2023biometric}
\bibinfo{author}{Pereira, T.M.}, \bibinfo{author}{Concei{\c{c}}{\~a}o, R.C.}, \bibinfo{author}{Sencadas, V.}, \bibinfo{author}{Sebasti{\~a}o, R.}, \bibinfo{year}{2023}.
\newblock \bibinfo{title}{Biometric recognition: A systematic review on electrocardiogram data acquisition methods}.
\newblock \bibinfo{journal}{Sensors} \bibinfo{volume}{23}, \bibinfo{pages}{1507}.
\bibitem[{Plataniotis et~al.(2006)Plataniotis, Hatzinakos and Lee}]{plataniotis2006ecg}
\bibinfo{author}{Plataniotis, K.N.}, \bibinfo{author}{Hatzinakos, D.}, \bibinfo{author}{Lee, J.K.}, \bibinfo{year}{2006}.
\newblock \bibinfo{title}{Ecg biometric recognition without fiducial detection}, in: \bibinfo{booktitle}{2006 Biometrics symposium: Special session on research at the biometric consortium conference}, \bibinfo{organization}{IEEE}. pp. \bibinfo{pages}{1--6}.
\bibitem[{Por{\'e}e et~al.(2016)Por{\'e}e, Kervio and Carrault}]{poree2016ecg}
\bibinfo{author}{Por{\'e}e, F.}, \bibinfo{author}{Kervio, G.}, \bibinfo{author}{Carrault, G.}, \bibinfo{year}{2016}.
\newblock \bibinfo{title}{Ecg biometric analysis in different physiological recording conditions}.
\newblock \bibinfo{journal}{Signal, image and video processing} \bibinfo{volume}{10}, \bibinfo{pages}{267--276}.
\bibitem[{Pu et~al.(2023)Pu, Wu, Wang, Sun, Liu and Liu}]{pu2023arrhythmia}
\bibinfo{author}{Pu, N.}, \bibinfo{author}{Wu, Z.}, \bibinfo{author}{Wang, A.}, \bibinfo{author}{Sun, H.}, \bibinfo{author}{Liu, Z.}, \bibinfo{author}{Liu, H.}, \bibinfo{year}{2023}.
\newblock \bibinfo{title}{Arrhythmia classifier based on ultra-lightweight binary neural network}, in: \bibinfo{booktitle}{2023 15th International Conference on Electronics, Computers and Artificial Intelligence (ECAI)}, \bibinfo{organization}{IEEE}. pp. \bibinfo{pages}{1--7}.
\bibitem[{Qiang et~al.(2024)Qiang, Dong, Liu, Yang, Hu and Wang}]{qiang2024ecgmamba}
\bibinfo{author}{Qiang, Y.}, \bibinfo{author}{Dong, X.}, \bibinfo{author}{Liu, X.}, \bibinfo{author}{Yang, Y.}, \bibinfo{author}{Hu, F.}, \bibinfo{author}{Wang, R.}, \bibinfo{year}{2024}.
\newblock \bibinfo{title}{Ecgmamba: Towards ecg classification with state space models}, in: \bibinfo{booktitle}{2024 IEEE International Conference on Bioinformatics and Biomedicine (BIBM)}, \bibinfo{organization}{IEEE}. pp. \bibinfo{pages}{6498--6505}.
\bibitem[{Schumann and B{\"a}r(2022)}]{schumann2022autonomic}
\bibinfo{author}{Schumann, A.}, \bibinfo{author}{B{\"a}r, K.J.}, \bibinfo{year}{2022}.
\newblock \bibinfo{title}{Autonomic aging--a dataset to quantify changes of cardiovascular autonomic function during healthy aging}.
\newblock \bibinfo{journal}{Scientific Data} \bibinfo{volume}{9}, \bibinfo{pages}{95}.
\bibitem[{Sumalatha et~al.(2024)Sumalatha, Prakasha, Prabhu and Nayak}]{sumalatha2024deep}
\bibinfo{author}{Sumalatha, U.}, \bibinfo{author}{Prakasha, K.K.}, \bibinfo{author}{Prabhu, S.}, \bibinfo{author}{Nayak, V.C.}, \bibinfo{year}{2024}.
\newblock \bibinfo{title}{Deep learning applications in ecg analysis and disease detection: An investigation study of recent advances}.
\newblock \bibinfo{journal}{IEEE Access} .
\bibitem[{Uwaechia and Ramli(2021)}]{uwaechia2021comprehensive}
\bibinfo{author}{Uwaechia, A.N.}, \bibinfo{author}{Ramli, D.A.}, \bibinfo{year}{2021}.
\newblock \bibinfo{title}{A comprehensive survey on ecg signals as new biometric modality for human authentication: Recent advances and future challenges}.
\newblock \bibinfo{journal}{IEEE Access} \bibinfo{volume}{9}, \bibinfo{pages}{97760--97802}.
\bibitem[{Wang et~al.(2024)Wang, Shanker, Nag, Lian and John}]{wang2024ecg}
\bibinfo{author}{Wang, G.}, \bibinfo{author}{Shanker, S.}, \bibinfo{author}{Nag, A.}, \bibinfo{author}{Lian, Y.}, \bibinfo{author}{John, D.}, \bibinfo{year}{2024}.
\newblock \bibinfo{title}{Ecg biometric authentication using self-supervised learning for iot edge sensors}.
\newblock \bibinfo{journal}{IEEE Journal of Biomedical and Health Informatics} .
\bibitem[{Wang et~al.(2007)Wang, Agrafioti, Hatzinakos and Plataniotis}]{wang2007analysis}
\bibinfo{author}{Wang, Y.}, \bibinfo{author}{Agrafioti, F.}, \bibinfo{author}{Hatzinakos, D.}, \bibinfo{author}{Plataniotis, K.N.}, \bibinfo{year}{2007}.
\newblock \bibinfo{title}{Analysis of human electrocardiogram for biometric recognition}.
\newblock \bibinfo{journal}{EURASIP journal on Advances in Signal Processing} \bibinfo{volume}{2008}, \bibinfo{pages}{1--11}.
\bibitem[{Wu et~al.(2020)Wu, Hung and Swindlehurst}]{wu2020ecg}
\bibinfo{author}{Wu, S.C.}, \bibinfo{author}{Hung, P.L.}, \bibinfo{author}{Swindlehurst, A.L.}, \bibinfo{year}{2020}.
\newblock \bibinfo{title}{Ecg biometric recognition: unlinkability, irreversibility, and security}.
\newblock \bibinfo{journal}{IEEE Internet of Things Journal} \bibinfo{volume}{8}, \bibinfo{pages}{487--500}.
\bibitem[{Wu et~al.(2021)Wu, Wei, Chang, Swindlehurst and Chiu}]{wu2021scalable}
\bibinfo{author}{Wu, S.C.}, \bibinfo{author}{Wei, S.Y.}, \bibinfo{author}{Chang, C.S.}, \bibinfo{author}{Swindlehurst, A.L.}, \bibinfo{author}{Chiu, J.K.}, \bibinfo{year}{2021}.
\newblock \bibinfo{title}{A scalable open-set ecg identification system based on compressed cnns}.
\newblock \bibinfo{journal}{IEEE Transactions on Neural Networks and Learning Systems} \bibinfo{volume}{34}, \bibinfo{pages}{4966--4980}.
\bibitem[{Yang et~al.(2018)Yang, Zhang, Yin and Liu}]{yang2018robust}
\bibinfo{author}{Yang, H.M.}, \bibinfo{author}{Zhang, X.Y.}, \bibinfo{author}{Yin, F.}, \bibinfo{author}{Liu, C.L.}, \bibinfo{year}{2018}.
\newblock \bibinfo{title}{Robust classification with convolutional prototype learning}, in: \bibinfo{booktitle}{Proceedings of the IEEE conference on computer vision and pattern recognition}, pp. \bibinfo{pages}{3474--3482}.
\bibitem[{Zhou et~al.(2024)Zhou, Huang, Liu, Zhang, Zhang and Chung}]{zhou2024open}
\bibinfo{author}{Zhou, S.}, \bibinfo{author}{Huang, X.}, \bibinfo{author}{Liu, N.}, \bibinfo{author}{Zhang, W.}, \bibinfo{author}{Zhang, Y.T.}, \bibinfo{author}{Chung, F.L.}, \bibinfo{year}{2024}.
\newblock \bibinfo{title}{Open-world electrocardiogram classification via domain knowledge-driven contrastive learning}.
\newblock \bibinfo{journal}{Neural Networks} \bibinfo{volume}{179}, \bibinfo{pages}{106551}.

\end{thebibliography}






\end{document}